\documentclass{svjour3a}

\usepackage{amsmath}
\usepackage{graphicx}
\setkeys{Gin}{draft=false}
\usepackage{natbib}

\usepackage{color}

\authorrunning{Arcos and LeVeque}

\titlerunning{Validating Velocities in the GeoClaw Tsunami Model}

\journalname{Pure \& Applied Geophysics}

\begin{document}
\title{Validating Velocities in the GeoClaw Tsunami Model using
Observations Near Hawaii from the 2011 Tohoku Tsunami}

\author{M. E. M. Arcos and Randall J. LeVeque}
\institute{%
     MEC, 180 Grand Ave, Suite 1100, Oakland, CA 94612
     \email{beth.arcos@amec.com} 
\and
    University of Washington \at
    Dept. Applied Mathematics 
    \email{rjl@uw.edu} }

\maketitle

\begin{abstract}
The ability to measure, predict, and compute
tsunami flow velocities is of importance in risk assessment and hazard
mitigation. Substantial damage can be done by high velocity flows, 
particularly in harbors and
bays, even when the wave height is small. Moreover, advancing
the study of sediment transport and tsunami deposits depends on the
accurate interpretation and modeling of tsunami flow velocities
and accelerations. Until recently, few direct measurements of tsunami velocities
existed to compare with model results. During the 11 March 2011 Tohoku
Tsunami 328 current meters were
in place around the Hawaiian Islands, USA, that captured time series of
water velocity in 18 locations, in both harbors and deep channels, at
a series of depths. 
We compare several of these velocity
records against numerical simulations performed using the GeoClaw
numerical tsunami model, based on solving the depth-averaged shallow water
equations with adaptive mesh refinement, to confirm that this model can
accurately predict velocities at nearshore locations. Model results
demonstrate tsunami current velocity is more spatially variable than
wave form or height and therefore may be a more sensitive variable for model validation.
\end{abstract}


\section{Introduction}\label{sec:intro}
In the last decade, tsunamis have resulted in millions of dollars in damage
far from their sources. Often this damage was not the result of high
flow depth or long inundation distances, but rather due to the strong currents generated
in harbors and along coastlines. In the last ten years, tsunamis
have generated over \$170 million damage in U.S. states and territories.
For example, the Mw 9.0 2011 Tohoku tsunami alone is estimated to have
caused \$90 million in damages in the U.S., even though the highest waves
arrived near low tide  and there was little on-shore inundation
(\cite{Western_states:2011}), and the Mw 8.3 2006
Kuril Island earthquake generated a tsunami that caused \$20 million
in damage to the Crescent City, California, harbor
(\cite{dengler-uslu:2011}). These costs highlight the need to
understand and predict the velocities of currents generated by
tsunamis in harbors and channels.
A recent study by \cite{lynett_assessment_2014}
illustrates that tsunami current velocities between 3 and 6 knots
(roughly 1.5 to 3 m/s) can cause moderate damage while velocities
above 6 knots can cause major damage.

The study of sediment transport and tsunami deposits also depends
on knowledge of flow velocities and accelerations (\cite{apotsos:2011}).
Determining potential for sediment transport and the source location of sediment transported
is of importance in establishing the frequency and magnitude of past
events based on the study of tsunami deposits
(\cite{Bourgeois:2009p888,Huntington:EOS}).  
Recent studies have focused on tsunami velocities and flow parameters
based on interpreting the source of sediment
(e.g., \cite{Moore:2007,Sawai:2009}). Accurate modeling
of locations and depths of strong flow and accelerations would aid
in determining possible sediment source locations.
Better understanding of tsunami current velocities may also be important in
exploring the depth at which tsunamis erode and deposit sediment, and whether
tsunamis can leave submarine records (\cite{Weiss:2008}).

Numerical tsunami models
are frequently validated using only wave height and inundation data,
which is now plentiful from recent events (e.g.,
\cite{liu:2005,synolakis:2005,satake:2005,macinnes:2009,macinnes:2013,okal:2010,fritz:2011}),
Surface elevation from Deep-Ocean Assessment and Reporting of
Tsunamis (DART) buoys or other deep-ocean monitors as well  as from
coastal and harbor tide gauges are used, along with inundation and
runup data collected by tsunami survey teams following every major
event (e.g., \cite{synolakis:2007,apotsos:2011}). 
See in particular the work of \cite{tang:2009}, which concerns
validation of the MOST tsunami model using data around Hawaii from several
past tsunamis.

The ability to measure and compute tsunami velocities remains at the
frontier of tsunami science. Tsunami models calculate depth-averaged
water velocity, but until recently there have been few data sets
available to validate the model results. Recent studies have begun to
compare tsunami velocity simulations with laboratory results,
post-tsunami survey data and analysis of survivor videos, or direct
measurement from current meters. Limited data are available of
directly measured tsunami currents. In the nearfield, tsunami flow
velocities have been calculated from video analysis of floating
objects  (up to 11 m/s) 
(\cite{Fritz:2006,Fritz:2012,EERI:2011}),
damage to structures (5-8 m/s) (\cite{EERI:2011}),
and sediment
transport (up to 14 m/s) \cite{apotsos:2011, Jaffe-Gelf:2007}.
The Maule, Chile tsunami was observed with velocities up to 0.36 m/s
in Monterey Bay, California (\cite{Lacy:2012}).
Current meter data in the farfield were also recorded during the
2011 Tohoku event in Humboldt Bay (0.36 -- 0.84 m/s;
\cite{Admire:2013, admire_observed_2014})
and in New Zealand (\cite{Lynett:2012}).  Some currents were observed 
in Hawaii during the 2006 Kuril Island event (\cite{Bricker:2007}).
Other studies have used video or ship GPS
analysis in farfield harbors (e.g., \cite{Admire:2013, admire_observed_2014, 
Lynett:2012}).

Tsunami modeling assumptions and validation can be tested with these
data. 
For example, \cite{apotsos:2011} compared numerical simulations to
both laboratory
experiments and field estimates of velocities during the 2004 Sumatra event.
In \cite{Jaffe-Gelf:2007,Fritz:2006,Fritz:2012,EERI:2011}, 
tsunami current
speeds interpreted from different tsunami events 
vary between 5 m/s and 14 m/s in the near field.
Direct measurements of tsunami velocity (up to 0.14 m/s) were obtained by current meters
deployed to monitor coral reef environments that serendipitously measured the
2006 Kuril Island tsunami (\cite{Bricker:2007}) at 40 depths in a water depth of 10 m
off the coast of Honolulu, Hawaii. These were used to test the
assumptions of shallow water wave equations with real world observations
(\cite{Arcas-Wei:2011}).
\cite{Lacy:2012}
observed the currents of the 2010 Maule, Chile tsunami at three depths
in Monterey Bay, California, recording velocities up to 0.36 m/s.
Higher velocities have been measured by analyzing survivor videos
of tsunami inundation near to the tsunami source. Velocities of up to 5 m/s and 11 m/s were measured of
the 2004 tsunami on the coast of Banda Aceh, Indonesia and the 2011 Tohoku
tsunami on the Sanriku coast of Japan respectively
(\cite{Fritz:2006,Fritz:2012}). In the Tohoku region of Japan, video
analysis esimated a flow velocity of 6.89 m/s during the 2011 tsunami 
(\cite{EERI:2011}). Based on flow depth and analysis of the properties
of building materials damaged in the 2011 tsunami on the Sendai Plain,
flow velocity estimates ranging from 5-8 m/s (\cite{EERI:2011}) were obtained. 
Using modeling of sedimentary data of
the 1998 Papua New Guinea tsunami a velocity of 14 m/s
was interpreted by \cite{Jaffe-Gelf:2007}. 
\cite{Admire:2013} and \cite{admire_observed_2014} studied currents induced
by the Tohoku event in 
Crescent City Harbor (estimated from video, up to 4.5 m/s) 
and Humboldt Bay (measured by a
current meter with a peak amplitude of 0.84 m/s), 
and compared with numerical simulations.
\cite{Lynett:2012} studied currents and vortical structures
in several ports and harbors induced by the Tohoku event.  

It is important to note that velocities often exhibit much greater
spatial variability than flow depth, particularly in bays and
harbors, and we present several figures to illustrate this.  This
may seem counter-intuitive for a model based on the shallow water
equations since for a long-wavelength coherent tsunami wave (e.g.
in the ocean away from shorelines) there is a direct relationship
between the surface elevation and the depth-averaged velocity, with
$s \approx \eta / \sqrt{h_0}$ where $s$ is the speed, $\eta$ the
surface displacement, and $h_0$ the undisturbed depth.  But this
is only true for a plane wave moving in one direction on a flat
bottom.  Anywhere there is a superposition of waves moving in
different direction (via reflections from shorelines and/or bathymetric
features) there is no longer such a clean relationship.
In shallow water near shore, the relationship will be even less clean
because of the interaction with bathymetry over short spatial scales 
on a much more finely resolved grid, and the generation of vorticity and
complex flows in harbors further complicate the picture.
This increases both the importance and difficulty of validating tsunami
models against observed velocities.

\begin{figure}
\hfil\includegraphics[width=\textwidth]{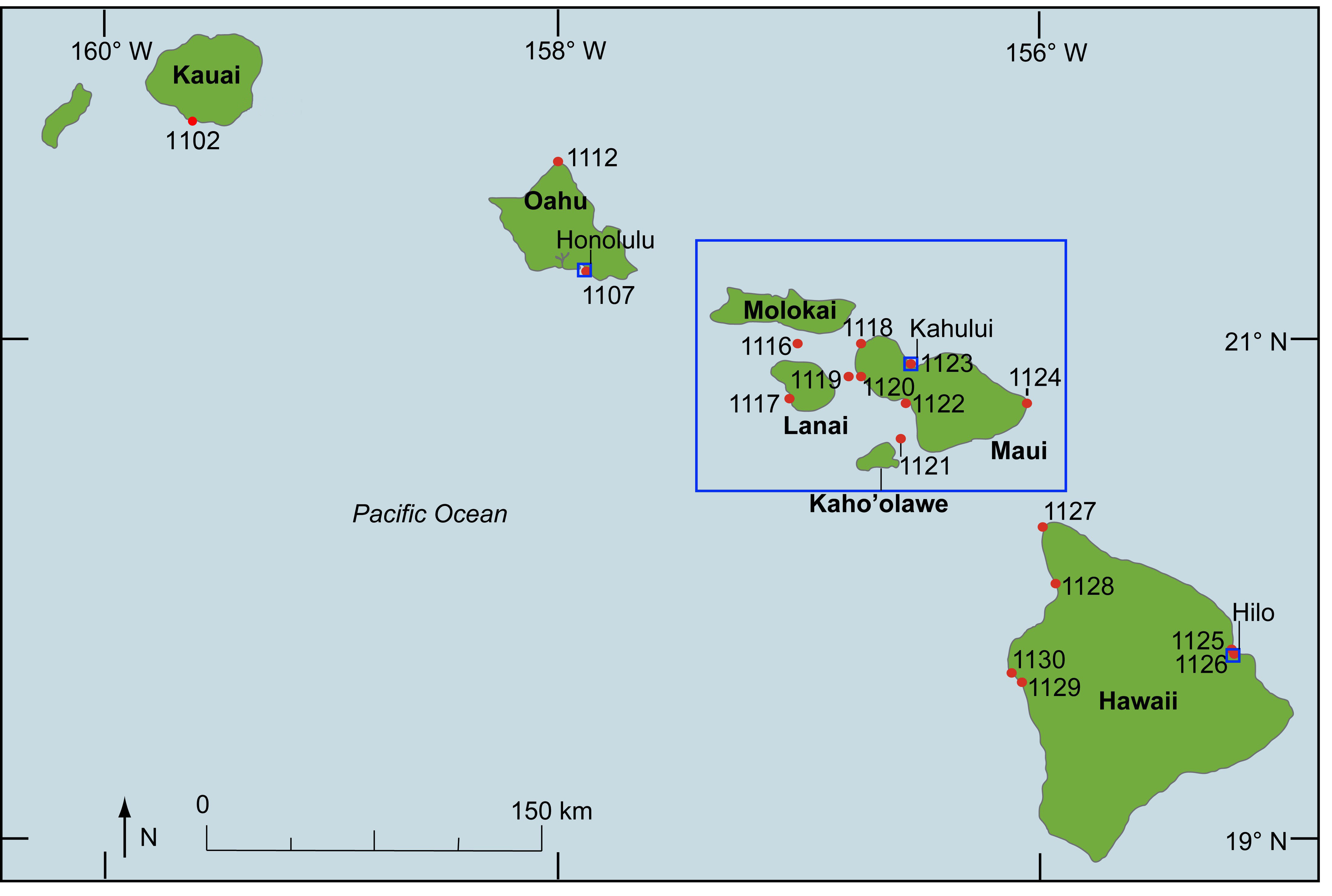}\hfil
\caption{\label{fig:map} 
Map showing the location of current velocity meter stations deployed around Hawaii,
including the ten stations used in this study. Blue boxes denote the
location of later maps.
}
\end{figure}

In this study, we explore the use of a data set from Hawaii for
the purpose of validating a numerical model.  During the 11 March
2011 Tohoku Tsunami, there were 328 current meters in place around
the islands of Hawaii that captured time series of the fluid velocity
at varying depths of water within the water column at 18 different
locations (stations) (\cite{coops-survey}) as shown in Figure~\ref{fig:map}.
Not only are these data direct measurements of the 2011
tsunami but they cover a much wider range of bathymetric conditions
than many previous studies including open coastline, deep channels and
harbors.
We have compared ten stations with simulations performed using the
GeoClaw software described in Section~\ref{sec:geoclaw}, which
solves the depth-averaged shallow water equations.  Section~\ref{sec:obs}
describes observations and simulation results. 
These validation results are presented in Section~\ref{sec:results}.
At most of these
stations the numerical model reproduces depth-averaged versions of
the measured data quite well, although the agreement is not as good at the
stations that are inside harbors, as discussed further in
Sections~\ref{sec:results} and \ref{sec:conclusions}.

The past studies that are most aligned with this paper are the work of 
\cite{Cheung:2013}, who also performed velocity
comparisons for all the stations used in this paper,
and \cite{Yamazaki:2012}, where velocity comparisons are made
at the Kilo Nalu Observatory near Honolulu Harbor.  In both case the
numerical simulations were performed using the NEOWAVE code.
The Kilo Nalu Observatory is indicated as KN in Figure~\ref{fig:Hon},
and discussed further in Section~\ref{sec:results}.

\section{The GeoClaw numerical model}\label{sec:geoclaw}

The open source GeoClaw tsunami model (\cite{geoclaw})
was used to perform numerical simulations.
This model has undergone extensive validation and verification tests
as reported in
\cite{rjl-george:catalina04,BergerGeorgeLeVequeMandli:awr11,%
geoclaw-nthmp-results:2011,%
LeVequeGeorgeBerger:an11}, using both
synthetic test problems and real events, but always based on comparing
surface elevations or inundation.  This paper presents the first
comparisons of GeoClaw results with current data and, as far as we
know, the first direct, quantitative comparison of observed time
series data from velocity meters to modeled results over such a broad
area and variety of settings. 

The GeoClaw software implements high-resolution finite volume methods to
solve the nonlinear shallow water equations, a depth-averaged system of
partial differential equations in which the fluid depth $h(x,y,t)$ and 
two horizontal depth-averaged velocities $u(x,y,t)$ and $v(x,y,t)$ are
introduced. $u$ is defined as the eastward and $v$ is
defined as the northward velocity component. These equations are written in a form that corresponds to
conservation of mass and momentum whenever the terms on the right hand side
vanish:
\begin{equation}\label{2dSWE}
\begin{split} 
  h_t + (hu)_x + (hv)_y &= 0,\\
  (hu)_t + \left(hu^2 + \frac 1 2 gh^2\right)_x + (huv)_y &= -ghB_x - Dhu,\\
  (hv)_t +  (huv)_x+ \left(hv^2 + \frac 1 2 gh^2\right)_y &= -ghB_y - Dhv,
\end{split} 
\end{equation}
Subscripts denote partial derivatives. The momentum source terms on the right hand side involve the varying
bathymetry $B(x,y,t)$ and a frictional drag term, where $D(h,u,v)$ is a
drag coefficient given by 
\begin{equation}\label{Dmanning}
D(h,u,v) = n^2gh^{-4/3}\sqrt{u^2+v^2}.
\end{equation}
The parameter $n$ is the {\em Manning coefficient} and depends on the
roughness.  If detailed information about the seafloor or inundated region
is known, then this
could be a spatially varying parameter. For generic tsunami modeling a
constant value of $n=0.025$ is often used.
The value $n=0.035$ has been suggested to better
account for the fringing reefs surrounding the Hawaiian
Islands by \cite{HawaiiManning}, 
and in this study we used this latter value, but also ran all of our
simulations with the smaller value and found virtually identical results.  
This is to be expected since the friction term generally makes a
difference only in very shallow water.  The choice of Manning coefficient
can make a significant difference for inundation studies, 
but not for the off-shore flow studied here.

Coriolis terms can also be added to the right hand side of equations
(\ref{2dSWE}), but these generally have been found to be negligible in tsunami
modeling (e.g., \cite{DaoTkalich:2007, KirbyShi:2013}).  
We have performed all of our computations both with and without
the Coriolis terms and have confirmed this for the case studied here ---
there is no visible difference in the gauge results when Coriolis terms are
added.

Other than the Manning friction coefficient, 
there are no tunable parameters in the GeoClaw model.  The sea level
parameter in the code can also be varied to adjust the initial water level.  
We have used the vertical datum of the bathymetry data, which for the
nearshore bathymetry is referenced to Mean High Water (MHW).
For modeling inundation, it may be important to more carefully set the tide
stage and Manning coefficient, but for the offshore gauges the differences
are negligible.

The finite volume methods implemented in GeoClaw are based on dividing the
computational domain into rectangular grid cells and storing cell averages
of mass and momentum in each grid cell. These are updated each time step by
a high-resolution Godunov type method (\cite{rjl:fvmhp}) that is based on
solving Riemann problems at the interfaces between neighboring grid cells
and applying nonlinear limiters to avoid nonphysical oscillations.  These
methods are second order accurate in space and time wherever the solution is
smooth, but robustly handle 
strong shock waves and other discontinuous solutions. This is 
important when the tsunami reaches shallow water and hydraulic jumps arise
from wave breaking.  
Also, the methods have been extended to deal robustly with inundation. Grid
cells where $h=0$ represent dry land and cells can dynamically change
between wet and dry each time step.

Block structured 
adaptive mesh refinement is used to employ much finer grid resolution in
regions of particular interest.  Regions of refinement track the tsunami as
it propagates across the ocean, and then additional levels of refinement are
added around Hawaii and even finer grids in the regions around the gauges of
interest.  For the calculations presented here, a grid resolution of
$2^\circ$ was used at the coarsest level, Five additional nested levels of
refinement were generally used, generally going down to $1''$ resolution on
the finest grid for gauges away from harbors (1 arcsecond is about 30 meters).
For simulations of Hilo and Kahului Harbors 
the finest level was reduced to $1/3''$, while for the larger region
surrounding Honolulu Harbor the finest level was $2/3''$.
Runs at finer resolution were used to confirm that the results shown here
are well resolved.

In addition to refining the spatial resolution,
smaller timesteps must be used on the finer grid patches. The GeoClaw
software implements the timestepping and transfer of information between
grids at different levels, as well as the automatic flagging of grid cells
requiring refinement and the clustering of these cells into rectangular
patches for refinement to the next level.
More details of these numerical methods can be found in
\cite{dgeorge:phd,dg-rjl:tsunami06,dgeorge:jcp,LeVequeGeorgeBerger:an11}.
Version 5.2.1 of the GeoClaw code was used to produce the final figures in this
publication.

The ETOPO1 global bathymetry (\cite{ETOPO1}) was used over the ocean, at 1 minute
resolution. 
Around the Hawaiian Islands bathymetry from the National
Geophysical Data Center was used, with resolution of $6''$, together
with finer $1/3''$ grids around Hilo and Honolulu Harbors (\cite{NGDC}). 
Near Kahului Harbor $1''$ resolution is the finest available and this was
used, and interpolated to $1/3''$ by the GeoClaw code.
GeoClaw combines data from different data sets
into a global piecewise bilinear function that can be integrated over each
computational grid cell in order to obtain the cell average of the
bathymetry.  This is done in a manner that is consistent between different
grid levels in order to maintain conservation of mass.

The initial seafloor deformation for the
tsunami simulations was based on \cite{fujii:2011}, 
which was calculated using tsunami waveform inversion based on DART buoy,
tidegauge and GPS gauge data.  This seafloor 
deformation was previously found to be one of the best at 
replicating nearfield run-up and DART buoy time series 
in a recent comparison of GeoClaw results using
ten proposed sources for the Tohoku event in the study of \cite{macinnes:2013}.

\section{Tsunami observations}\label{sec:obs}

In order to evaluate shipping lanes and harbors, the National Oceanic and
Atmospheric Administration deployed 30 current meter stations with
horizontal 2-D Acoustic Doppler Profilers (Sontek/YSIH - ADP)  off the
Hawaiian Islands in early 2011 as part of the 2011 Hawaii Current
Observation Project (\cite{coops-survey}). Eighteen stations were active and
recorded the passing of the Tohoku tsunami.  The station locations are
tabulated with additional information in Table~1
and are shown in Figure~\ref{fig:map}, and in more detail near the harbors in
Figures~\ref{fig:Hon} through \ref{fig:Hilo}.

Water depth at the active stations varied between 12.5 and 153 m. Current
meters were approximately evenly spaced at varying depths along a cable, with
6 to 34 current meters at each station with an increasing number of
current meters with increasing depth. The deepest current meters were between 1.6 and 29 m from the
seafloor and the shallowest were 2 to 17 m below the sea surface. The
current meters record speed and horizontal direction at six-minute time
intervals. The six-minute sample interval captures variation in
velocity with the inflow and outflow of long period tsunami waves but
may not capture other types of waves such as edge waves (periods of
three or more  minutes) that are excited by the tsunami and have been
observed with more detailed sample intervals
(\cite{Bricker:2007,Cheung:2013}).
Accuracy of the current meters is $\pm 0.5$ cm/s speed and $\pm 2$ degrees for direction. 
Wave amplitude and vertical motion were not recorded (\cite{coops-survey}). Vertical flow in
offshore tsunami waves is generally 
negligible compared to horizontal flow (\cite{Arcas-Wei:2011}). 

In this paper we present results for ten of these stations that
test the ability of our numerical model
to reproduce the primary characteristics of tsunami currents in
different settings.  
We focus on stations that are in protected water, primarily in the
channels between the 
islands of Maui, Molokai, and Lanai and near the harbors of Honolulu,
Kahului, and Hilo.  At these stations there is a strong tsunami signal visible
in the record for many hours after the arrival of the first wave, typically
because of seiching in these protected regions. These are also the regions
where tsunami currents are of most interest in relation to hazards to
harbors and shipping.  The stations studied include four in quite
shallow water, less than 25 m depth, near the harbors and bays (HAI1107,
HAI1123, HAI1125, and HAI1126), and six others in
in the inter-island channels 
(HAI1116, HAI1118, HAI1119, HAI1120, HAI1121, HAI1122).
With the exception of HAI1120 (near Lahaina), these are in 25--150 m depth.

\begin{figure}
\hfil\includegraphics[width=0.48\textwidth]{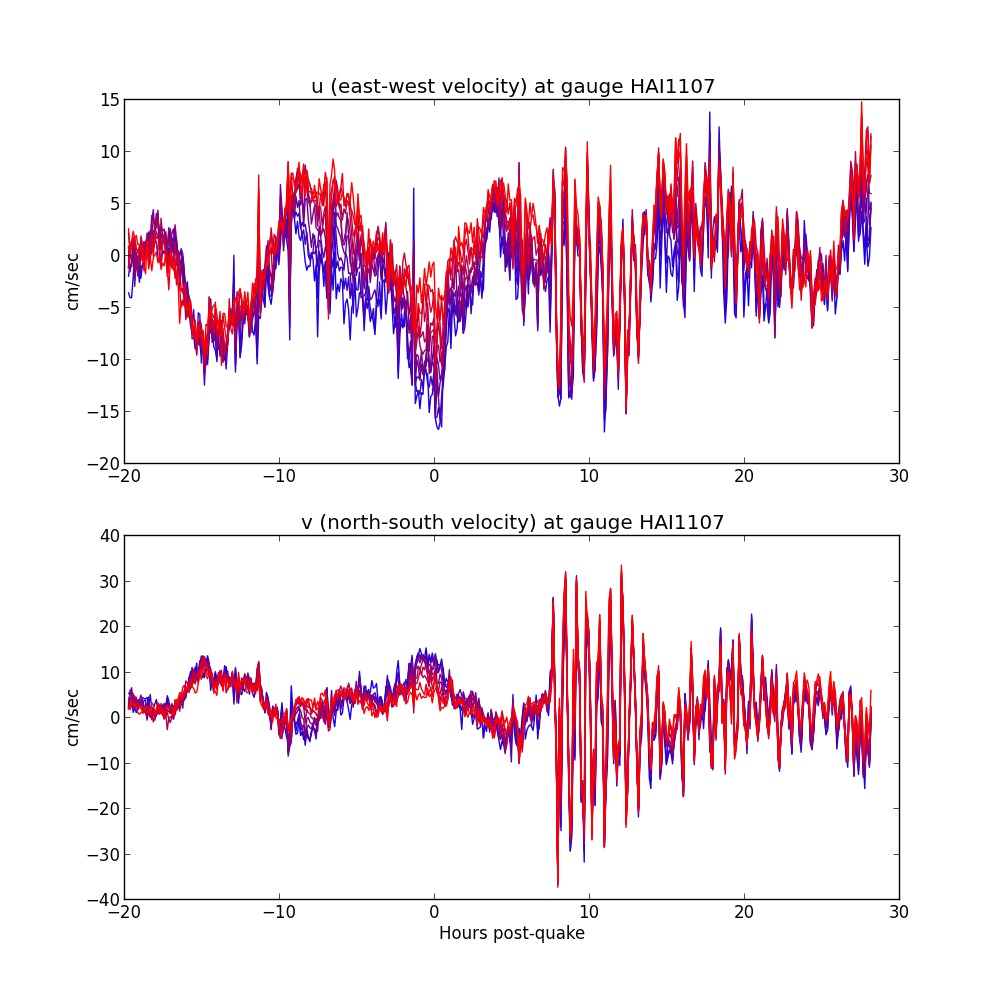}
\hfil\includegraphics[width=0.48\textwidth]{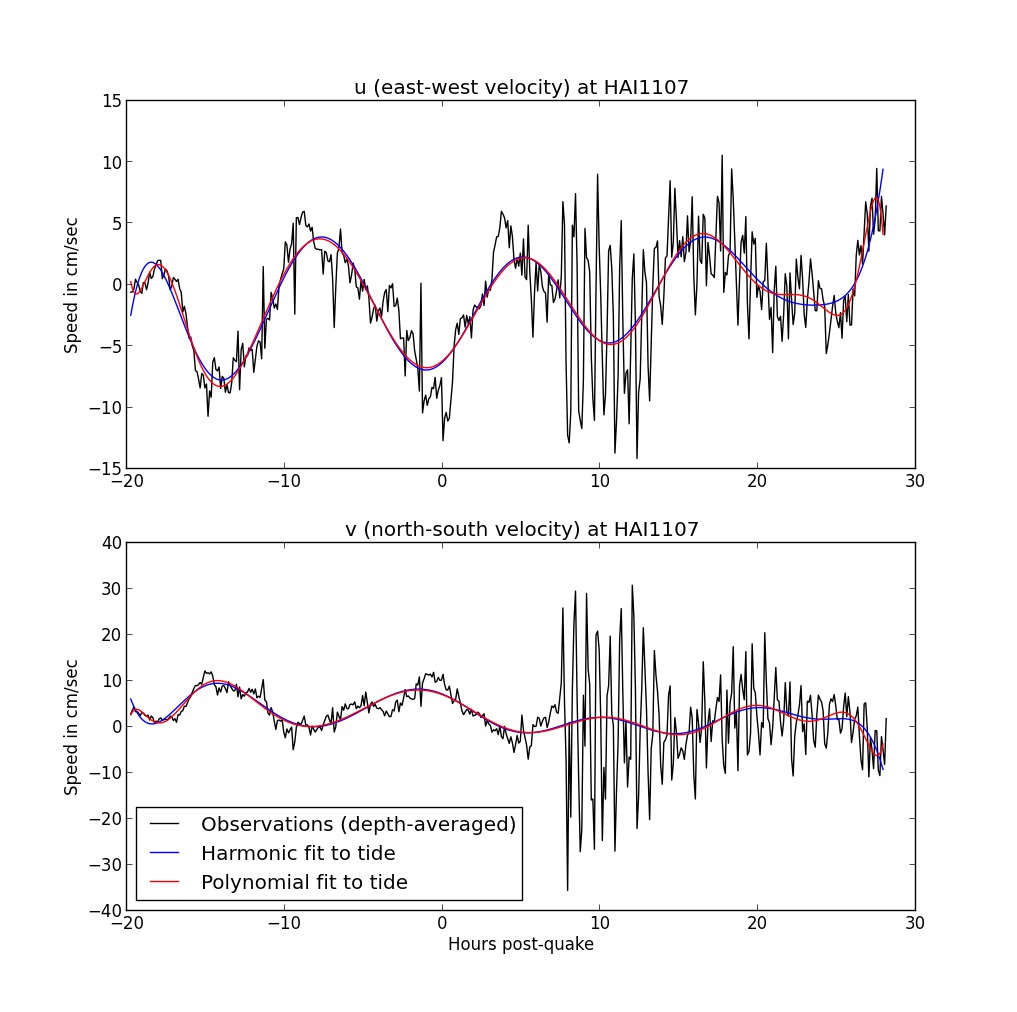}
\vskip 5pt
\hfil\includegraphics[width=0.48\textwidth]{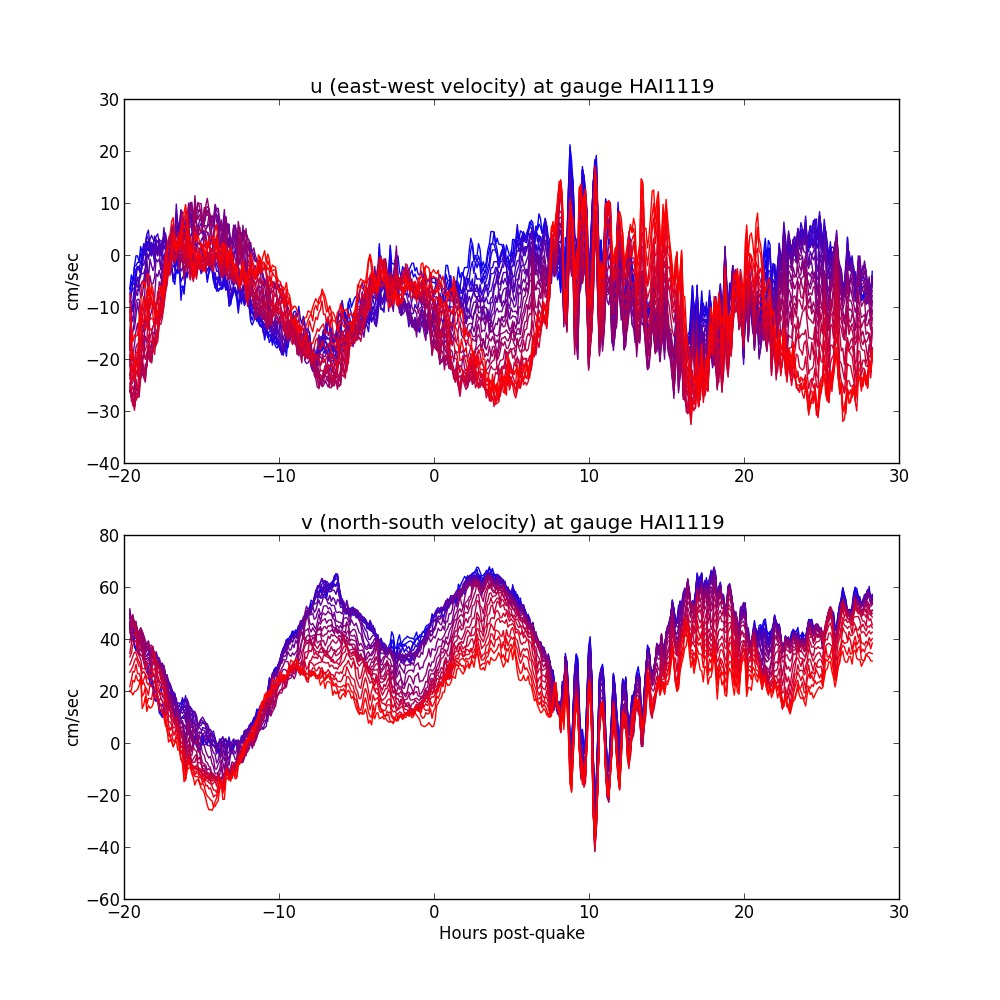}
\hfil\includegraphics[width=0.48\textwidth]{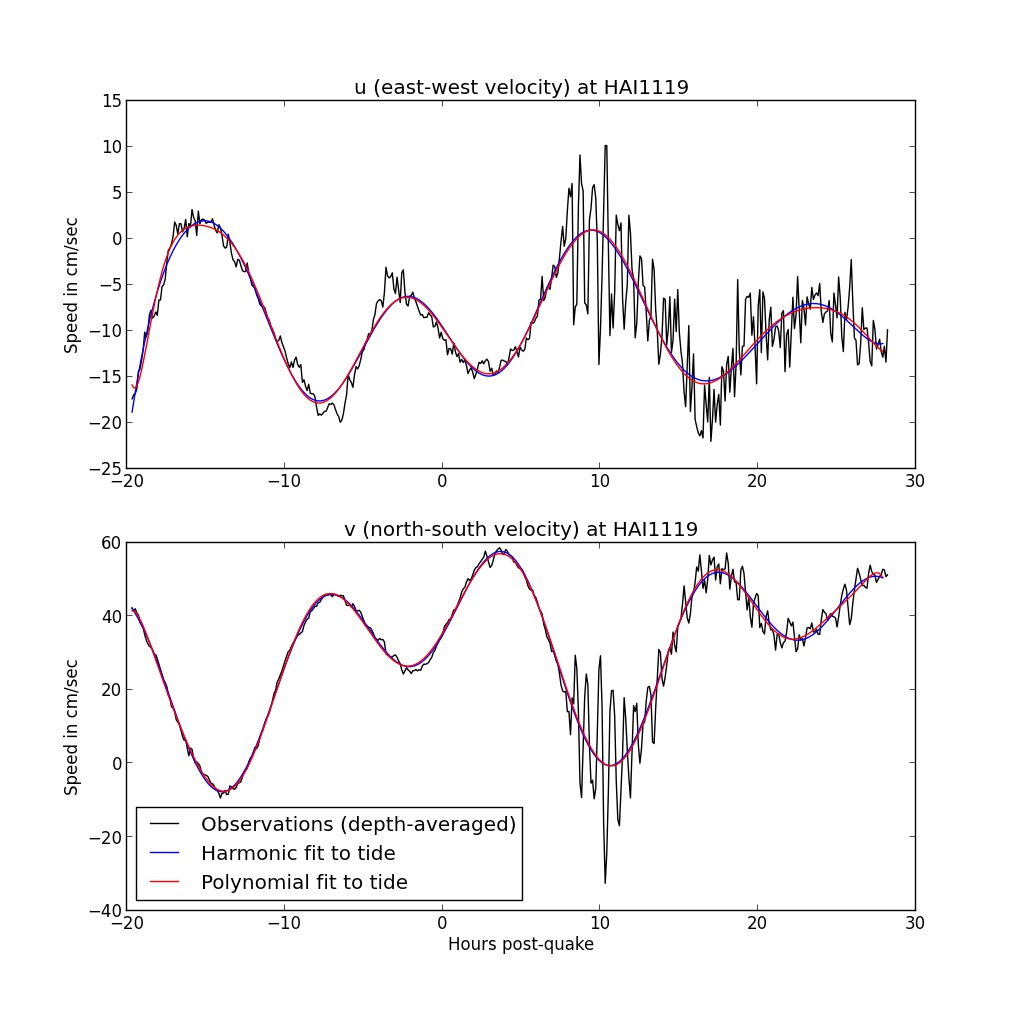}
\caption{\label{fig:rawdata} 
Plots of observed velocities at two sample stations, HAI1107, Honolulu
Harbor, and HAI1119, Auau Channel.  On the left, velocities are shown at all
depths, split into u and v velocity components. Blue is the shallower
gauges, Red is the deeper gauges. 
On the right, the depth averaged velocities are shown in black and the tidal
component based on harmonic constituents is shown in blue.  Also shown in
red is a polynomial fit to the tidal component.  In subsequent
figures the tidal component as determined by the harmonic fit
 is subtracted from the data.
Shown over a 48-hour window around the tsunami arrival time at roughly 8
hours post-quake.
}
\end{figure}

From the speed and direction of flow that was recorded at each
depth, we computed the east-west velocity $u$ and north-south velocity
$v$.  Two samples of these records are shown in Figure~\ref{fig:rawdata},
for a 48-hour window around the tsunami arrival time.
The top of Figure~\ref{fig:rawdata} shows HAI1107, in the approach to Honolulu
Harbor.  In this shallow location (14.9 m) there is very little
variation in velocity with depth.  Figure~\ref{fig:rawdata} also shows
HAI1119,  at 73.51 m depth in the Auau Channel, where the greatest
variation of velocity with depth was observed.  

Like all other codes for modeling transoceanic tsunamis, the
GeoClaw software computes a single
depth-averaged velocity at each point
and cannot directly model the velocity profile with depth.  
For comparison purposes, we depth-averaged the observed $u$ and $v$
velocities at each
depth to obtain a single velocity time series at each station.
This depth-averaged velocity is also shown in the right panel of
Figure~\ref{fig:rawdata}.
Note that the current meters generally did not span the entire water
column from seafloor to surface.  Hence we are averaging over only a subset
of the water column.  Since the shallow water equations assume a constant
velocity with depth, we believe this gives the best value for comparison
with the numerical results.

Current meter data has been filtered to remove tidal currents.  
Because the data appears so noisy, even before the arrival of the tsunami,
we explored two different approaches to detiding the data and found that they
gave nearly identical results, increasing our confidence in the results.
In both cases a least squares fit to a 
48-hour time series of data starting 20 hours before the earthquake was
computed.  
One approach was to fit a high degree polynomial, and we found that a
degree 20 polynomial was able to match the tidal oscillations in this length
data without introducing higher frequency oscillations, and that results
were fairly insensitive to the degree.  The second
approach was to use a more traditional harmonic constituent approach, where
the data is fit by a sum of sines and cosines with periods given by the
10 dominant tidal constituents in this region.  
We found that this results in a poorly conditioned least squares problem and
so we used the singular value decomposition to compute a regularized
solution by discarding components corresponding to
singular values smaller than $10^{-5}$ times the maximum singular value.

The plots on the right of Figure~\ref{fig:rawdata} also show these two curve
fits for each velocity component at each of the two sample stations.
The two fits lie virtually on top of each other, particularly in the region
from 7 to 13 hours after the earthquake, when the first tsunami waves arrive
in Hawaii.  All subsequent plots focus on this time period and the harmonic
constituent fit to the tide has been subtracted from the raw data, both for
the velocity gauge data and for tide gauge data.

All GeoClaw time series results in the figures below
have been uniformly shifted by adding 10 minutes 
to the time from the computation.  This was done because all computational
results showed approximately the same phase shift relative to the
observations and performing this shift makes it much easier to assess the
accuracy of the amplitude and period relative to the observational results.
Possible reasons for this phase shift are discussed in
Section~\ref{sec:arrival}.

\section{Results}\label{sec:results}

\begin{figure}
\hfil\includegraphics[width=0.45\textwidth]{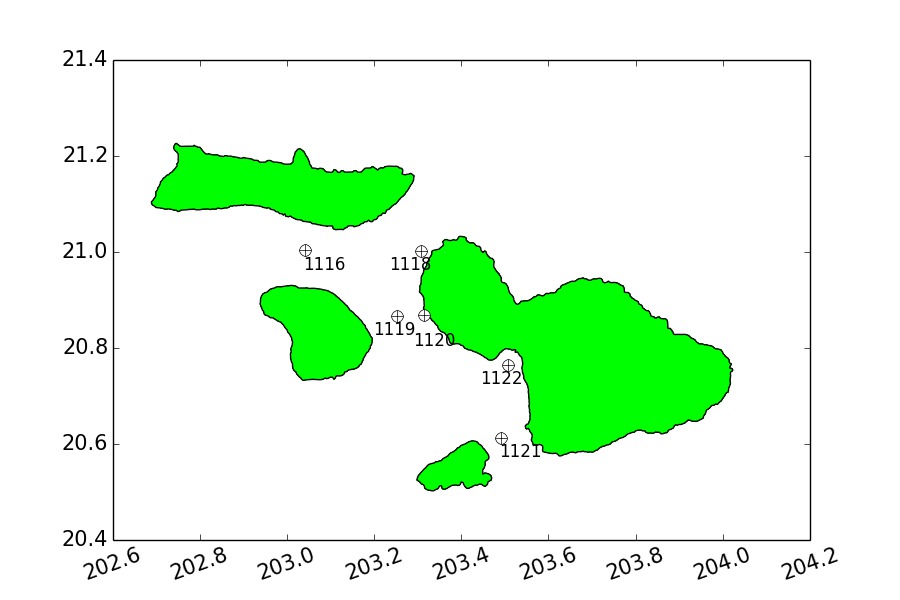}\hfil
\hfil\includegraphics[width=0.55\textwidth]{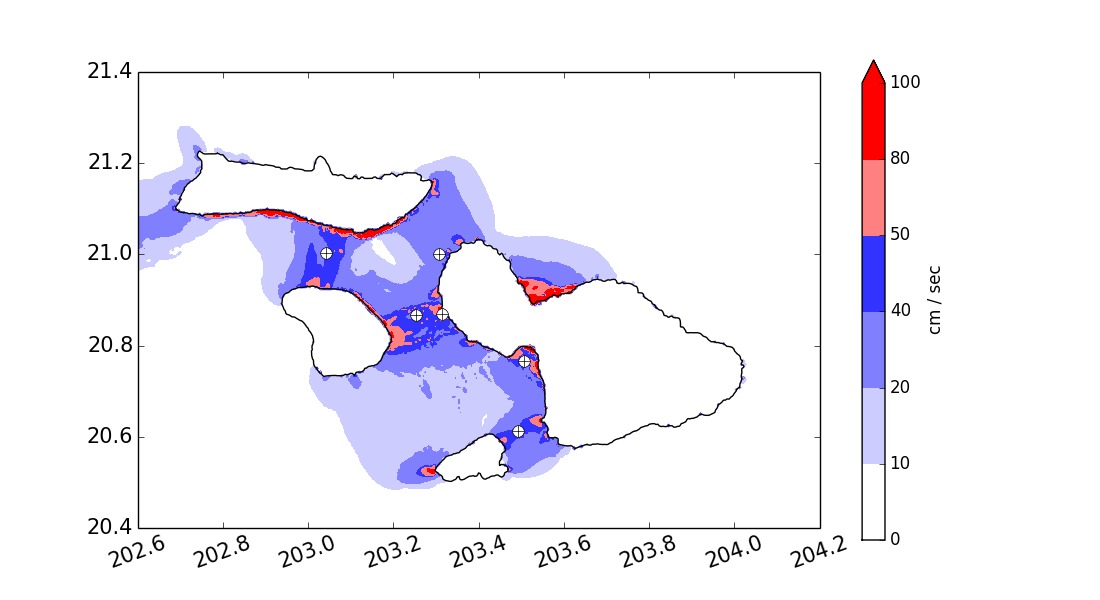}\hfil
\caption{\label{fig:channel} 
Left: Station locations in the inter-island channels (boxed region 
of Figure~\ref{fig:map}).
Right: Maximum flow speed from model simulation (scale in cm/s).
}
\end{figure}

The left panel of Figure~\ref{fig:channel} shows the boxed region from 
Figure~\ref{fig:map} and the stations that lie in the inter-island channels
between the islands of Maui, Molokai, and Lanai.  
The simulations reveal that there is much greater spatial
variation in flow speed than is typically observed in sea surface elevation.
This makes it potentially more challenging to accurately compute the flow
velocity at any particular point.  To illustrate this variation,
the right panel of
Figure~\ref{fig:channel} shows the maximum computed flow speed
$s=\sqrt{u^2 + v^2}$
calculated over the entire 13 hours of simulated time within this region.
Note that velocities are much larger between the
islands than in the surrounding waters (where tsunami currents are generally
less than 10 cm/s), and are largest in the narrowest
constrictions between islands, up to 50 cm/s. 
This is consistent with what is expected
from the fluid dynamics and observed in other tsunamis (e.g.
\cite{borrero:1997,Lynett:2012}).

\begin{figure}
\hfil\includegraphics[width=0.5\textwidth]{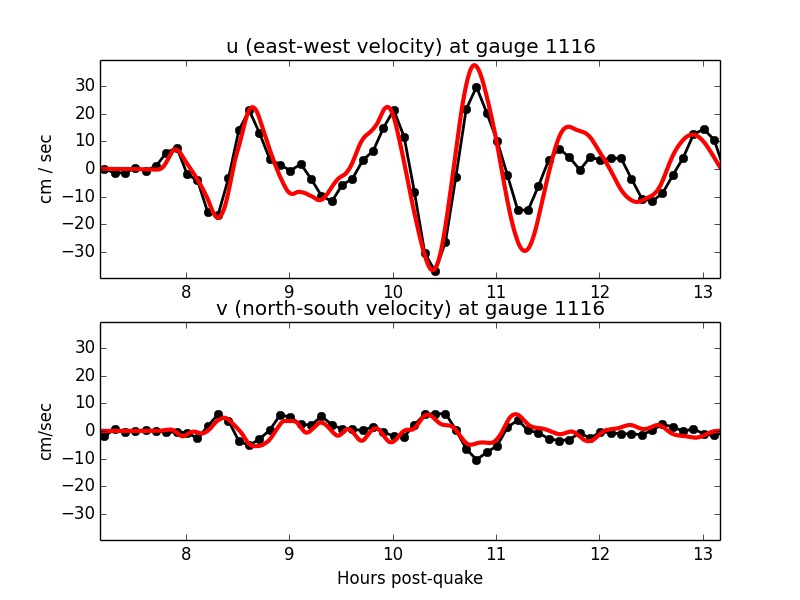}\hfil
\hfil\includegraphics[width=0.5\textwidth]{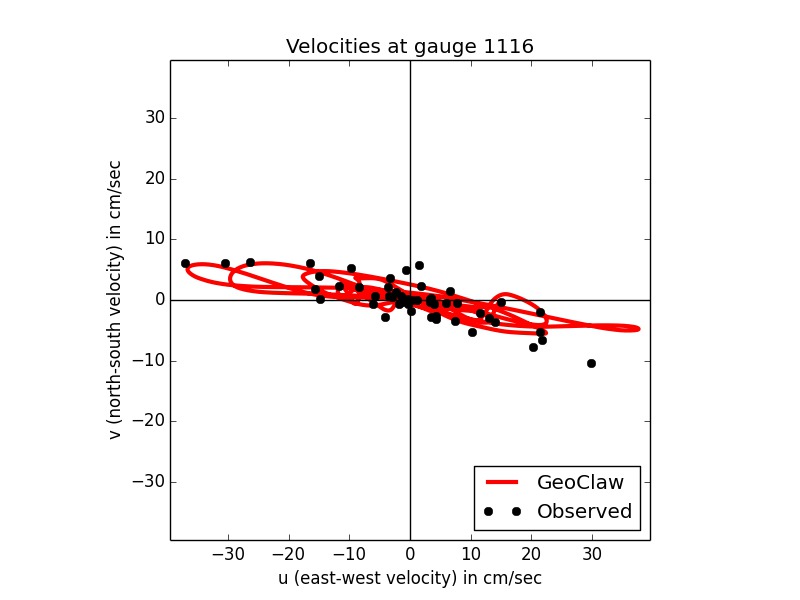}\hfil
\vskip 10pt
\hfil\includegraphics[width=0.5\textwidth]{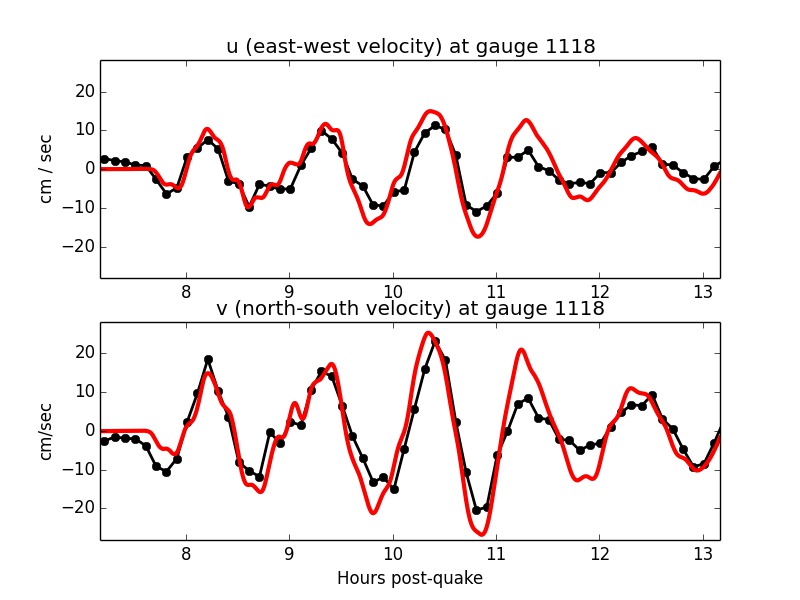}\hfil
\hfil\includegraphics[width=0.5\textwidth]{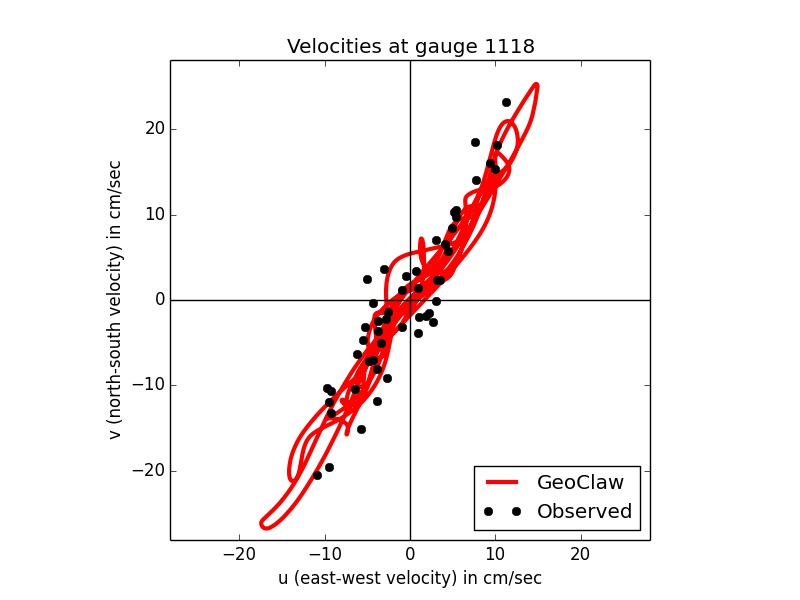}\hfil
\vskip 10pt
\hfil\includegraphics[width=0.5\textwidth]{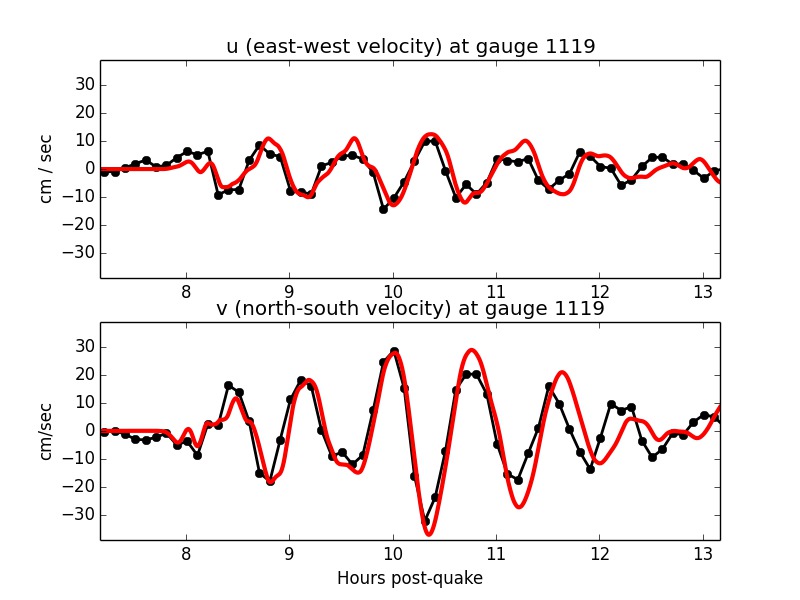}\hfil
\hfil\includegraphics[width=0.5\textwidth]{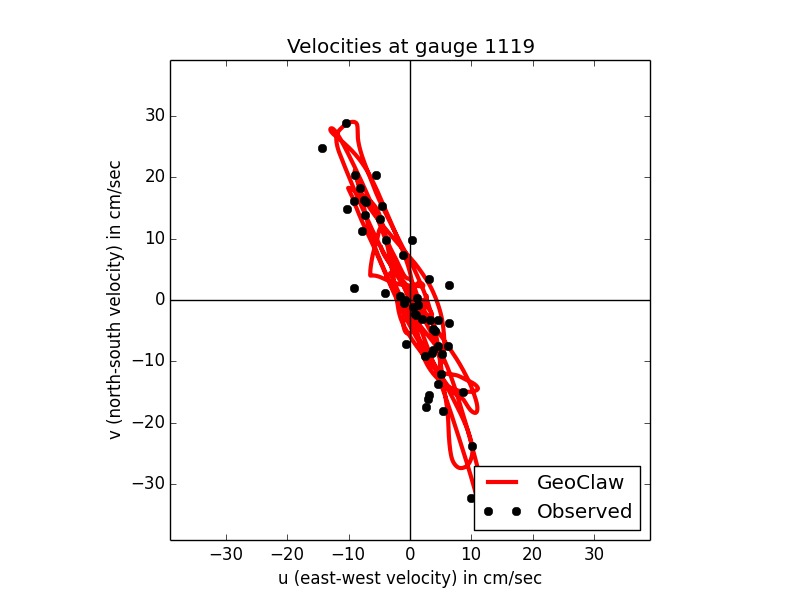}\hfil
\vskip 10pt
\caption{\label{fig:channels1} 
Observed depth-averaged velocity compared to simulated
tsunami velocity at each current meter over a 6 hour window at stations
HAI1116, 1118, and 1119.  
Left: $u$ and $v$ components of velocity vs.\ time.
The GeoClaw results (red solid line) have been shifted by +10 minutes in all
cases as discussed in the text.
Right: Plotted in the $u$--$v$ plane, showing direction of flow.
}
\end{figure}

\begin{figure}
\hfil\includegraphics[width=0.5\textwidth]{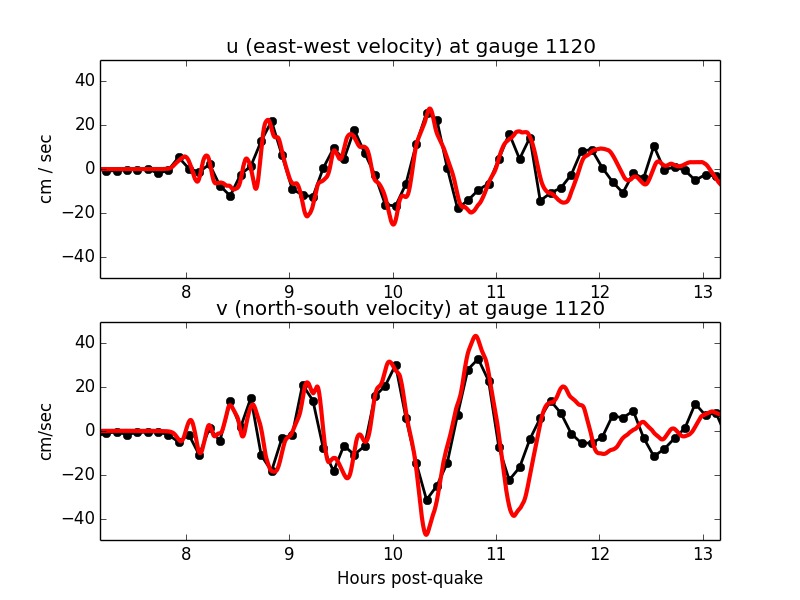}\hfil
\hfil\includegraphics[width=0.5\textwidth]{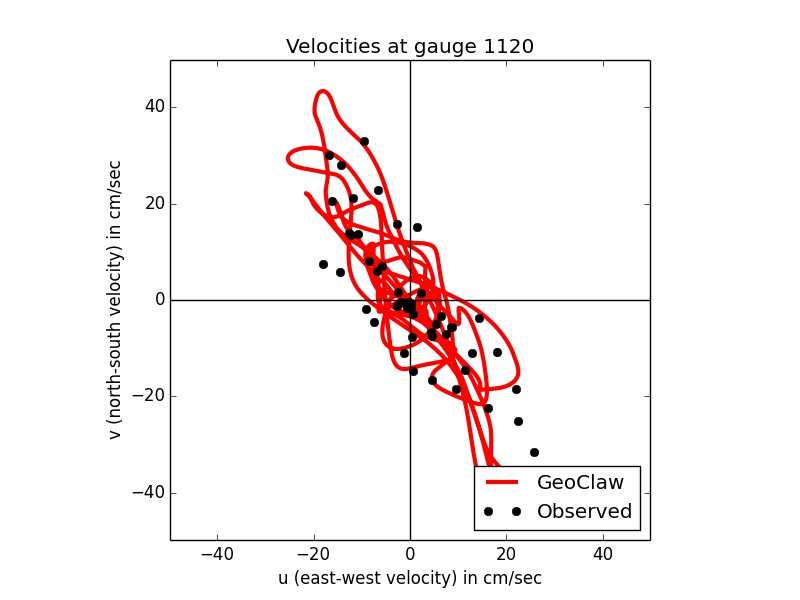}
\vskip 10pt
\hfil\includegraphics[width=0.5\textwidth]{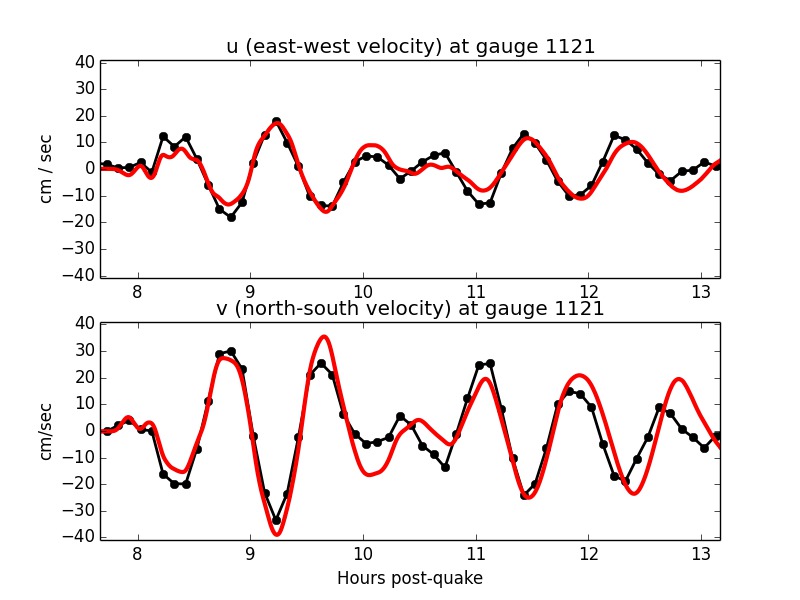}\hfil
\hfil\includegraphics[width=0.5\textwidth]{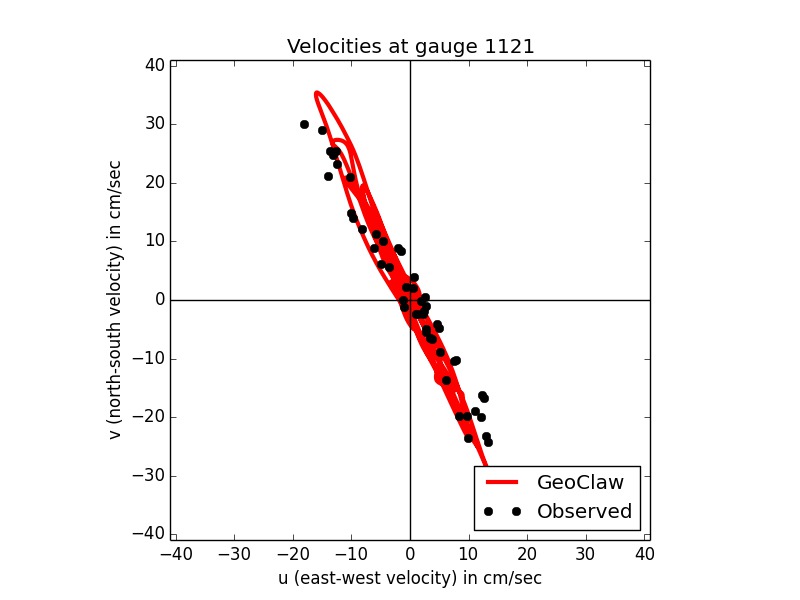}\hfil
\vskip 10pt
\hfil\includegraphics[width=0.5\textwidth]{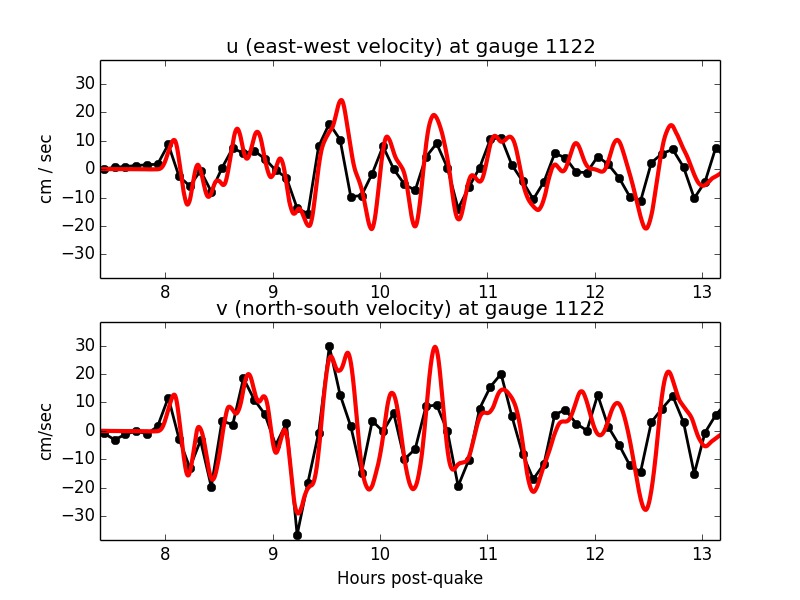}\hfil
\hfil\includegraphics[width=0.5\textwidth]{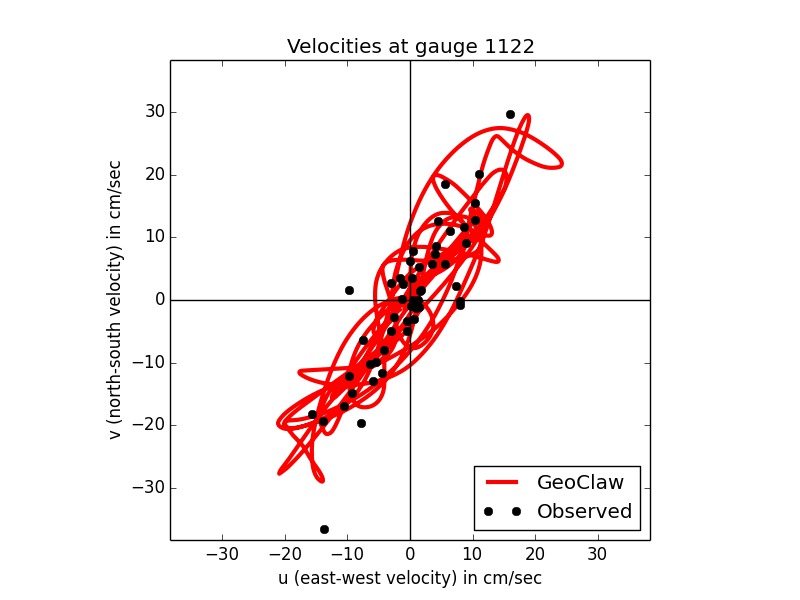}\hfil
\vskip 10pt
\caption{\label{fig:channels2} 
Observed depth-averaged velocity compared to simulated
tsunami velocity at each current meter over a 6 hour window at stations
HAI1120, 1121, and 1122.  
See Figure~\ref{fig:channels1} caption for description of plots.
plots.
}
\end{figure}

Figures  \ref{fig:channels1} and \ref{fig:channels2} show the computed
flow velocities at the stations in these channels, plotted along
with the observations in two different forms.  In the left column,
the east-west velocity $u$ and north-south velocity $v$ are 
plotted as time series for roughly 6 hours
after the tsunami arrival time.  In the right column of 
each figure we plot both the observed and computed
velocities in the $u$--$v$ plane to show the direction of flow.  
In general, the computed flow direction matches the observed direction
quite well, along with the amplitudes.
The periods and general wave form are also very similar at almost all these
inter-island stations.

\begin{figure}
\hfil\includegraphics[width=0.5\textwidth]{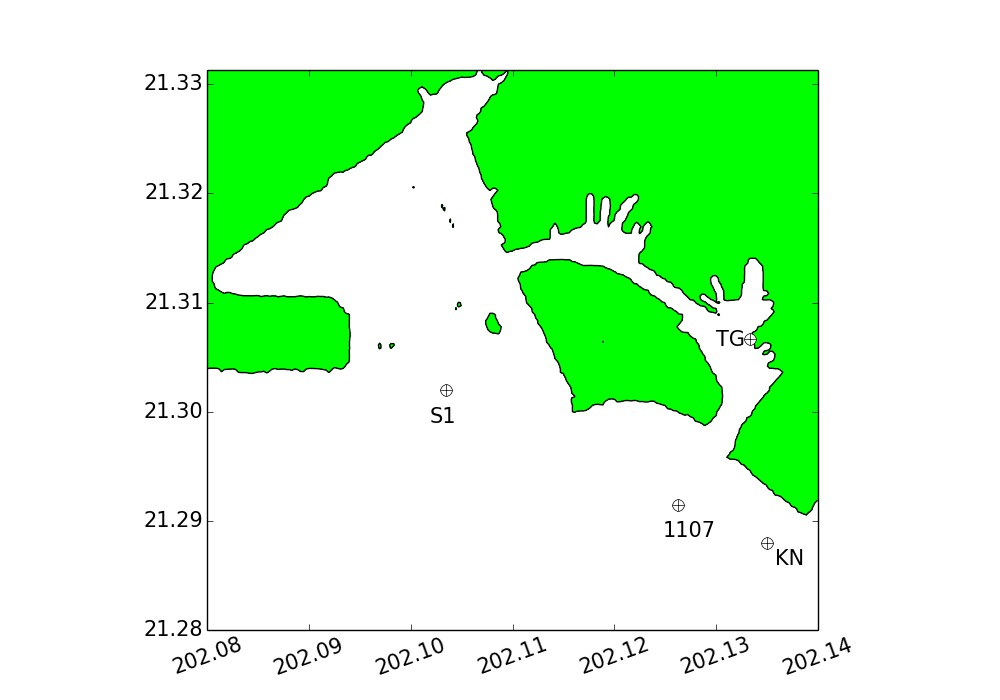}\hfil
\hfil\includegraphics[width=0.5\textwidth]{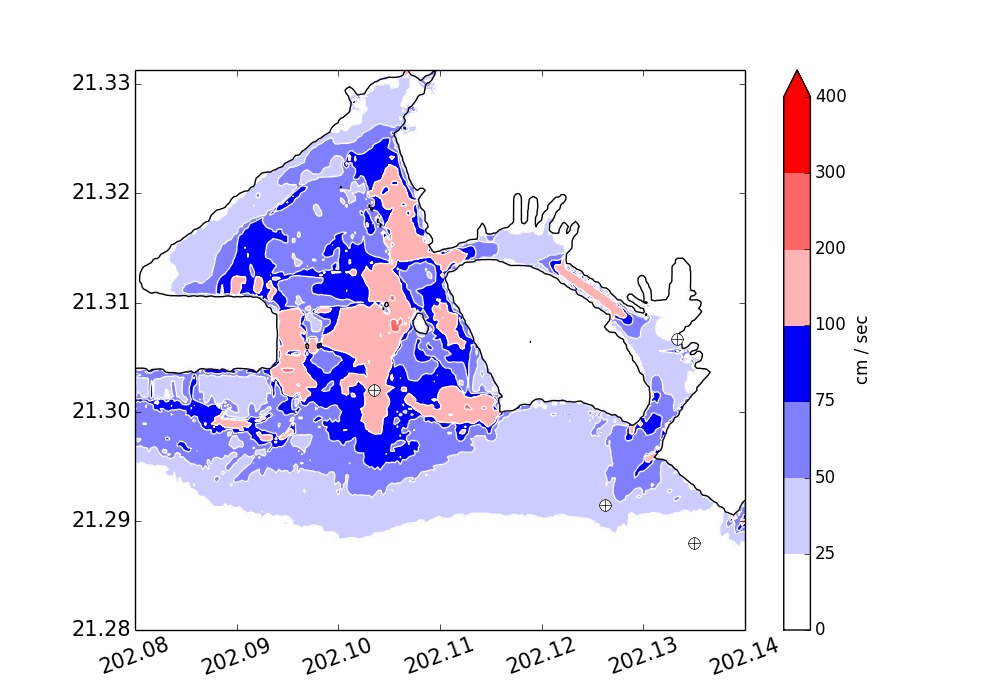}\hfil
\caption{\label{fig:Hon} 
Left: Station locations near Honolulu Harbor, including HAI1107,
tide gauge 1612340
(TG), a synthetic gauge (S1), and the Kilo Nalu Observatory (KN). 
Right: Maximum flow speed from model simulation (scale in cm/s).
}
\end{figure}

\begin{figure}
\hfil\includegraphics[width=0.5\textwidth]{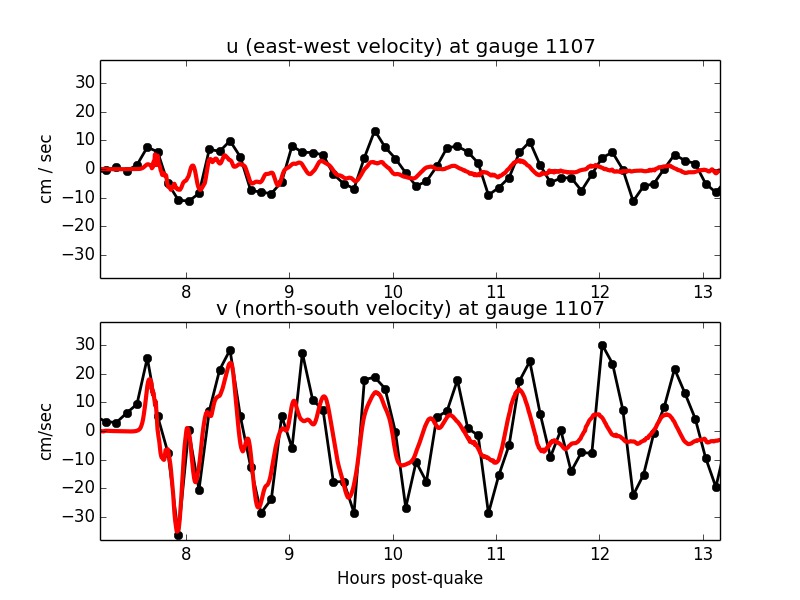}\hfil
\hfil\includegraphics[width=0.5\textwidth]{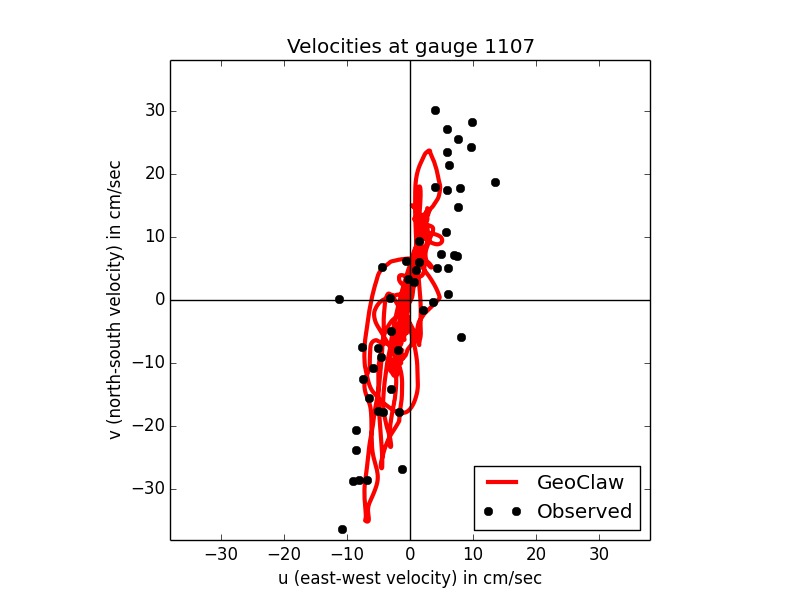}\hfil
\caption{\label{fig:HAI1107} 
Observed depth averaged velocity compared to simulated
tsunami velocity at each current meter over a 6 hour window at station HAI1107.
See Figure~\ref{fig:channels1} caption for description of plots.
}
\end{figure}

We also considered four stations that are in or near harbors: 
HAI1107 (Honolulu), HAI1123 (Kahului), and HAI1125, HAI1126 (Hilo).
The left panel of Figure~\ref{fig:Hon} shows the location of gauge HAI1107
near Honolulu Harbor, along with the tide gauge (TG) where surface elevation
data is available.  
The right panel of Figure~\ref{fig:Hon} shows the
maximum speed that was observed over the full simulation at each point in
this region. 

Figure  \ref{fig:HAI1107} shows the computed
flow velocities at HAI1107, plotted in the same manner as
Figures~\ref{fig:channels1} and \ref{fig:channels2}.
Again the results agree reasonably well in amplitude, phase and direction,
particularly for the first waves.  

\begin{figure}
\hfil\includegraphics[width=0.33\textwidth]{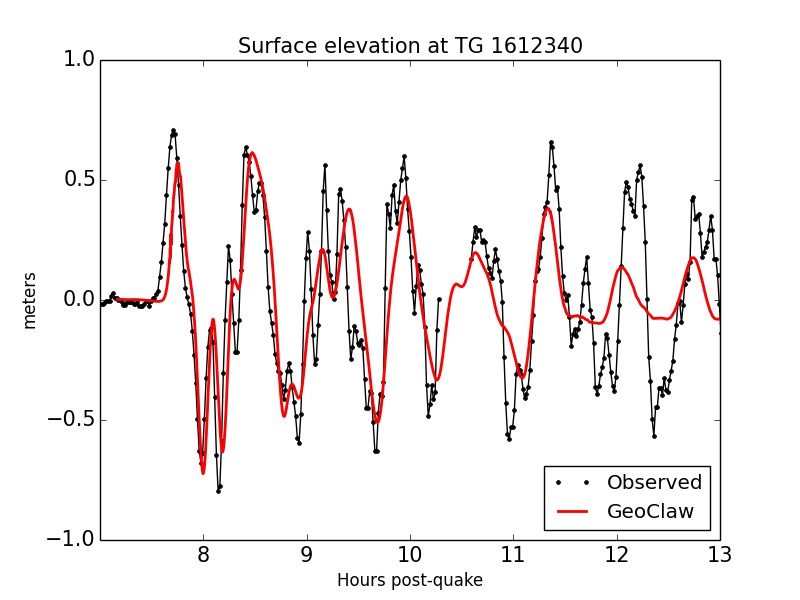}\hfil
\hfil\includegraphics[width=0.33\textwidth]{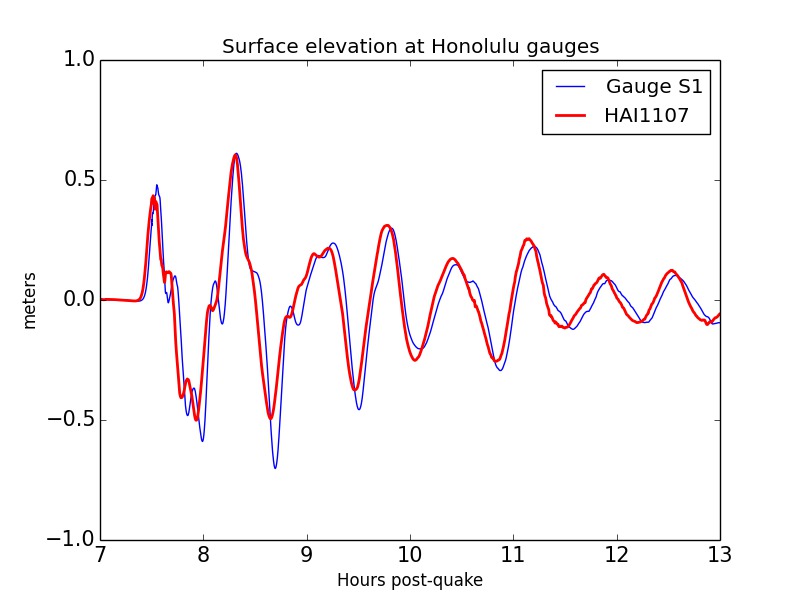}\hfil
\hfil\includegraphics[width=0.33\textwidth]{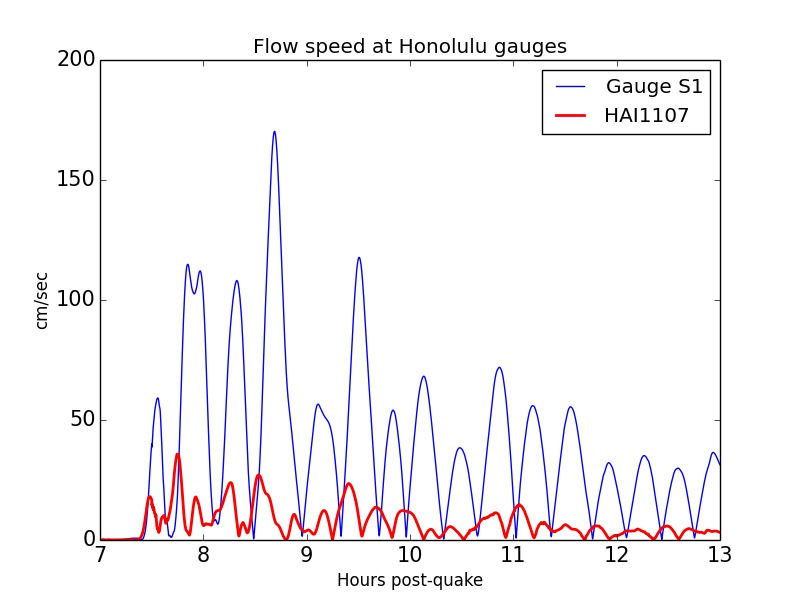}\hfil
\caption{\label{fig:HonGauges} 
Left: Observed and computed surface elevation at Tide gauge 1612340 Honolulu,
Center: Computed surface elevation at HAI1107 and a nearby synthetic gauge S1,
Right: Computed flow speed at HAI1107 and a nearby synthetic gauge S1.
This figure illustrates that there is much greater spatial variation 
in velocities than elevation.  See Figure~\ref{fig:Hon} for station locations.
}
\end{figure}

The left panel of Figure~\ref{fig:HonGauges} shows the
measured sea surface elevation (after de-tiding as described above) at the
tide gauge 1612340, along with the GeoClaw simulation results at the same point.
Such comparisons are typical of the
manner in which numerical tsunami models have been validated in the past
against tide gauge data.  
The amplitudes and periods match quite well between the observations and
computations, at least for the first several waves.

The remaining plots in Figure~\ref{fig:HonGauges} further illustrate
the fact that velocities can exhibit much greater spatial variation
than surface elevation.  The middle panel of Figure~\ref{fig:HonGauges}
shows computed surface elevations at station HAI1107 and an additional
synthetic station marked S1 in Figure~\ref{fig:Hon} that was used
in the numerical simulation, located in a dredged ship channel (no
observations are available at this point).  The time series of
surface elevation is virtually identical between these gauges, and
they are also similar to the surface elevations at the tide gauge
further back in the harbor.  On the other hand, the right panel of
Figure~\ref{fig:HonGauges} shows the time series of computed speeds
$s=\sqrt{u^2 + v^2}$ at the two locations HAI1107 and S1, and it
is seen that the speed varies by more than a factor of 10 and has
quite a different temporal pattern.  In view of this, we find it
particularly notable that the numerical model is able to calculate
results that agree as well as they do with observations of velocity
at the particular points where the gauges were located.

Figure~\ref{fig:Hon} also shows the location of the Kilo Nalu Observatory
marked as KN.  Velocity data at a single depth of 12 m at this location
are presented in the work of 
\cite{Yamazaki:2012}, and Figure 3 from that paper
shows a maximum flow speed of roughly 25 cm/sec. This is roughly
consistent with our figure for the maximum depth-averaged velocity, 
which shows that this meter was in a region of
even lower maximum current velocity than HAI1107.

\begin{figure}
\hfil\includegraphics[width=0.47\textwidth]{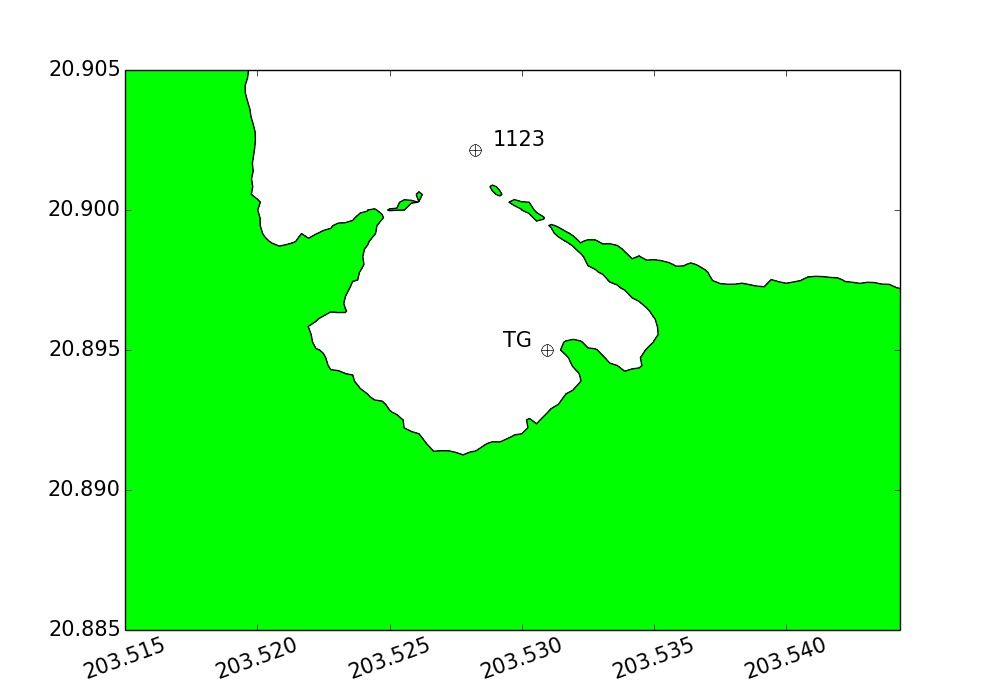}\hfil
\hfil\includegraphics[width=0.53\textwidth]{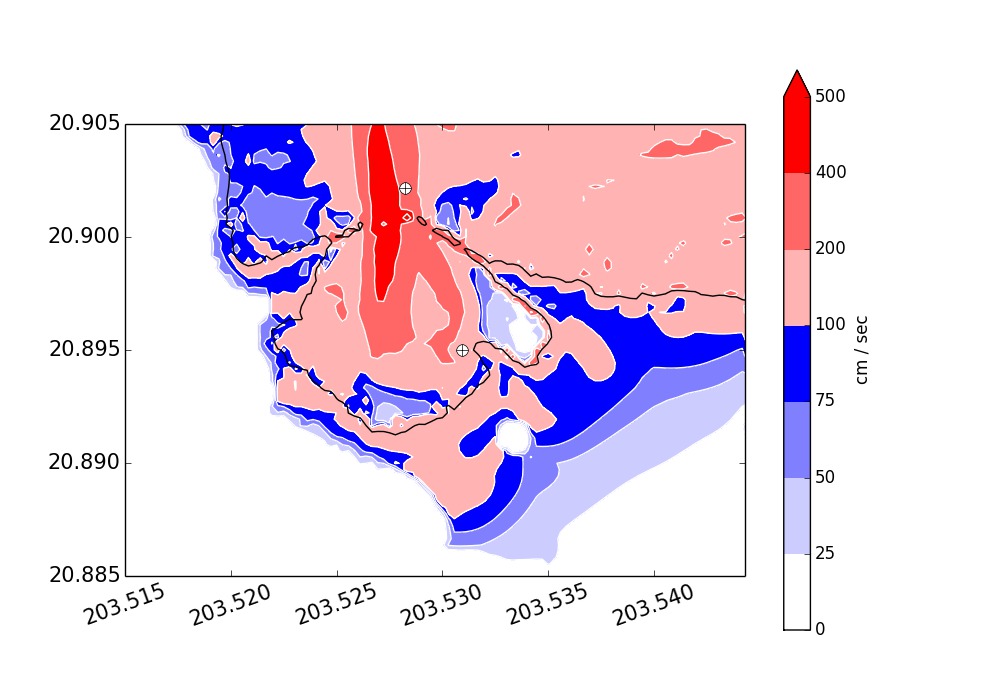}\hfil
\caption{\label{fig:Kahului} 
Left: Station locations near Kahului Harbor, including tide gauge 1615680 (TG). 
Right: Maximum flow speed from model simulation (scale in cm/s). 
}
\end{figure}

\begin{figure}
\hfil\includegraphics[width=0.5\textwidth]{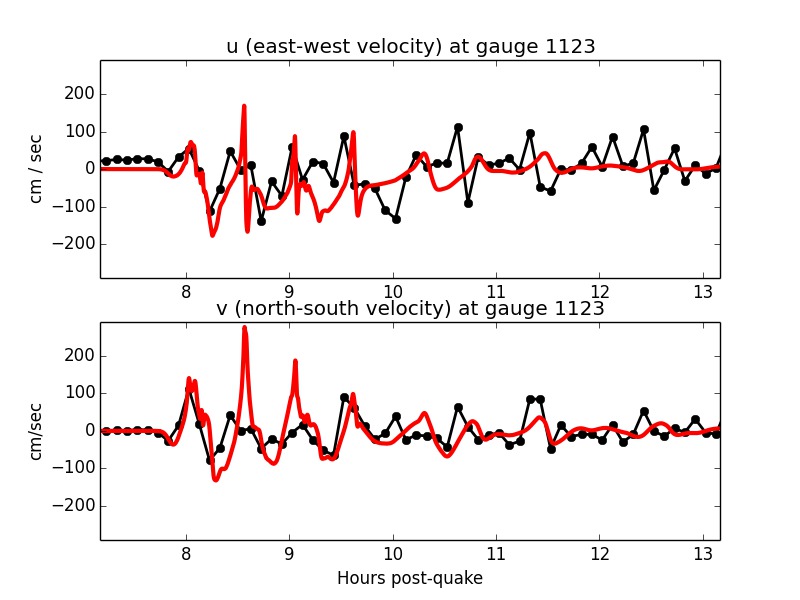}\hfil
\hfil\includegraphics[width=0.5\textwidth]{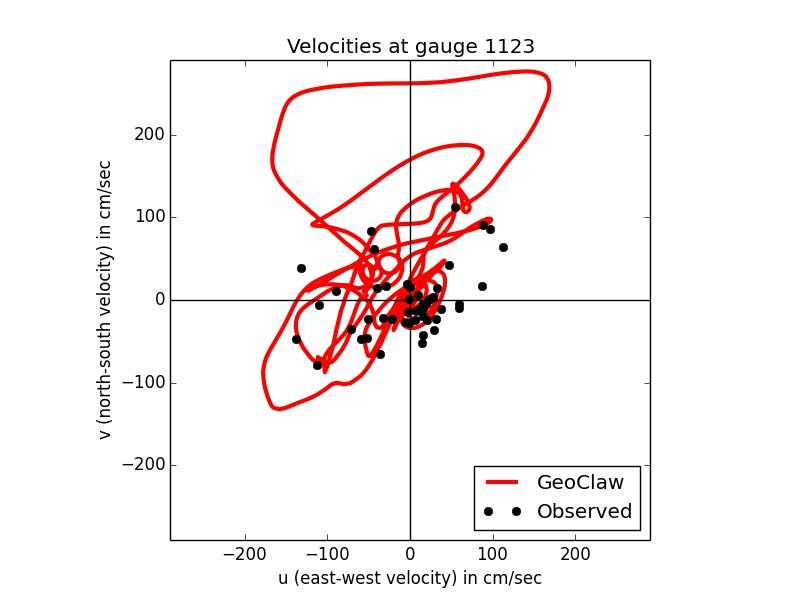}\hfil
\caption{\label{fig:HAI1123} 
Observed depth-averaged velocity compared to simulated
tsunami velocity at each current meter over a 3 hour window at station
HAI1123.  
See Figure~\ref{fig:channels1} caption for description of plots.
}
\end{figure}

\begin{figure}
\hfil\includegraphics[width=0.5\textwidth]{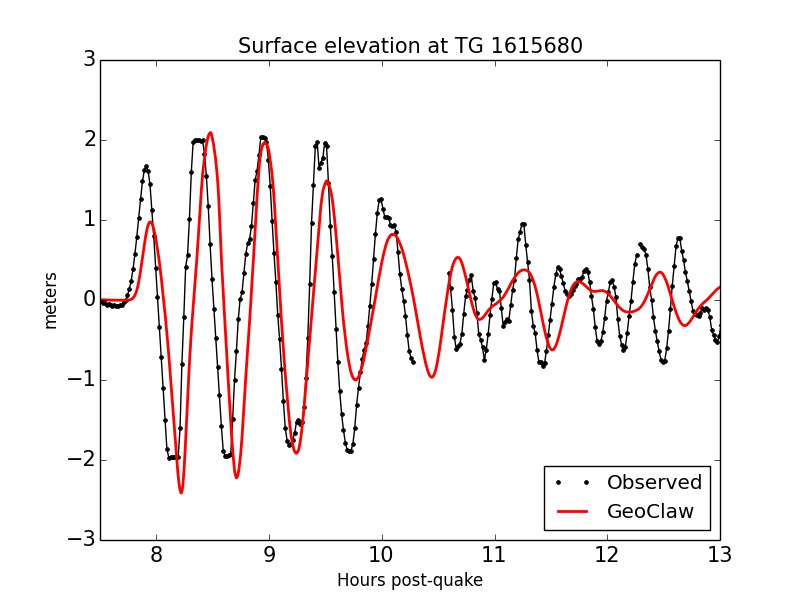}\hfil
\hfil\includegraphics[width=0.5\textwidth]{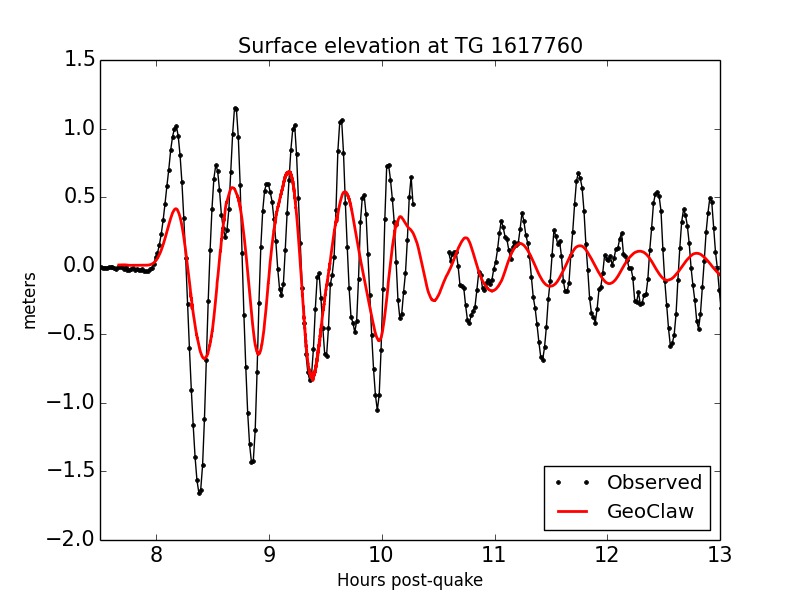}\hfil
\caption{\label{fig:TG} 
Surface elevation at tide gauges  1615680 Kahului (left) and
1617760 Hilo (right).  
The GeoClaw results (red solid line) have been shifted by +10 minutes in all
cases as discussed in the text.
}
\end{figure}

Station HAI1123 is 
very close to the entrance to Kahului Harbor. The location is shown in
Figure~\ref{fig:Kahului}, which also illustrates how much spatial
variation there is in maximum velocity near this harbor due to the 
bathymetry.  For this harbor only $1''$ bathymetry data are available (unlike
Honolulu and Hilo Harbors, where $1/3''$ bathymetry data have been used). 
The observations also show a lack of clear directionality, indicating that
the flow near the harbor entrance may have been turbulent.
In view of these considerations, 
the relatively poor agreement at HAI1123  in Figure~\ref{fig:HAI1123}
is perhaps not surprising.
But note also from Figure~\ref{fig:Kahului} that the 
maximum velocity is very sensitive to the exact position of the gauge and 
shifting it slightly to the east would give smaller amplitude velocities
that might better match the observations.  
This extreme spatial sensitivity may make it impossible to achieve close
agreement, even if the model were perfect, since the location of this gauge
is not precisely known.  
Three digits to the right of the decimal are recorded in the station metadata.
Even if all these digits
are correct, an uncertainty of $0.0005^\circ$ is roughly 50 meters.
Figure~\ref{fig:TG} shows the surface elevation at the tide gauge 1615680 in
Kahului Harbor, which shows better agreement than the velocity results in
this case.

\begin{figure}
\hfil\includegraphics[width=0.5\textwidth]{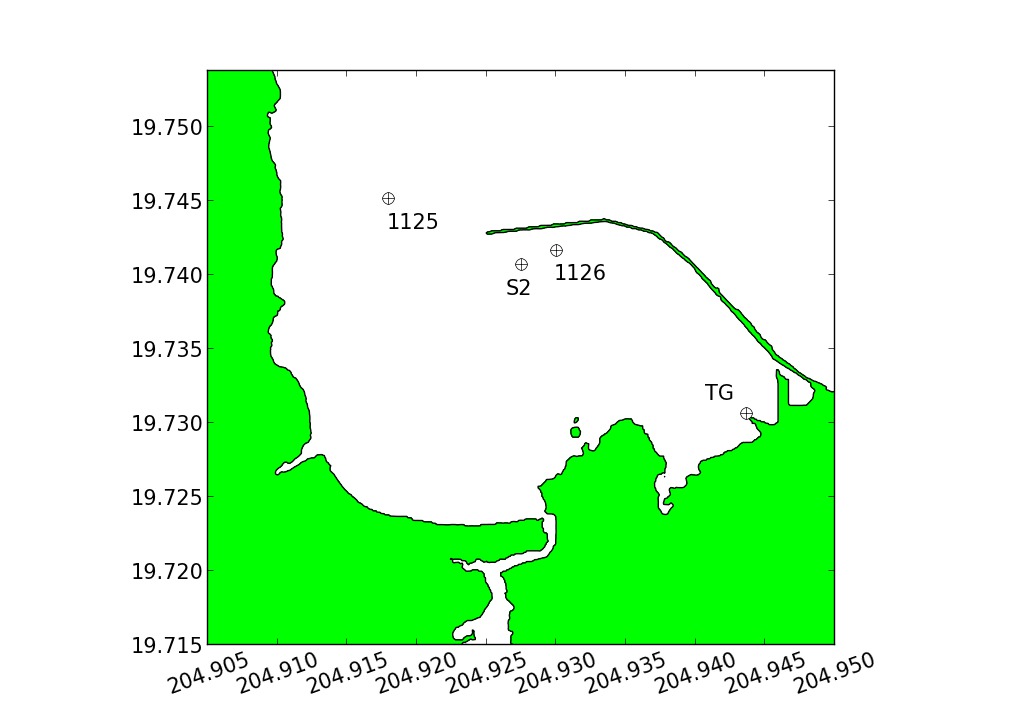}\hfil
\hfil\includegraphics[width=0.5\textwidth]{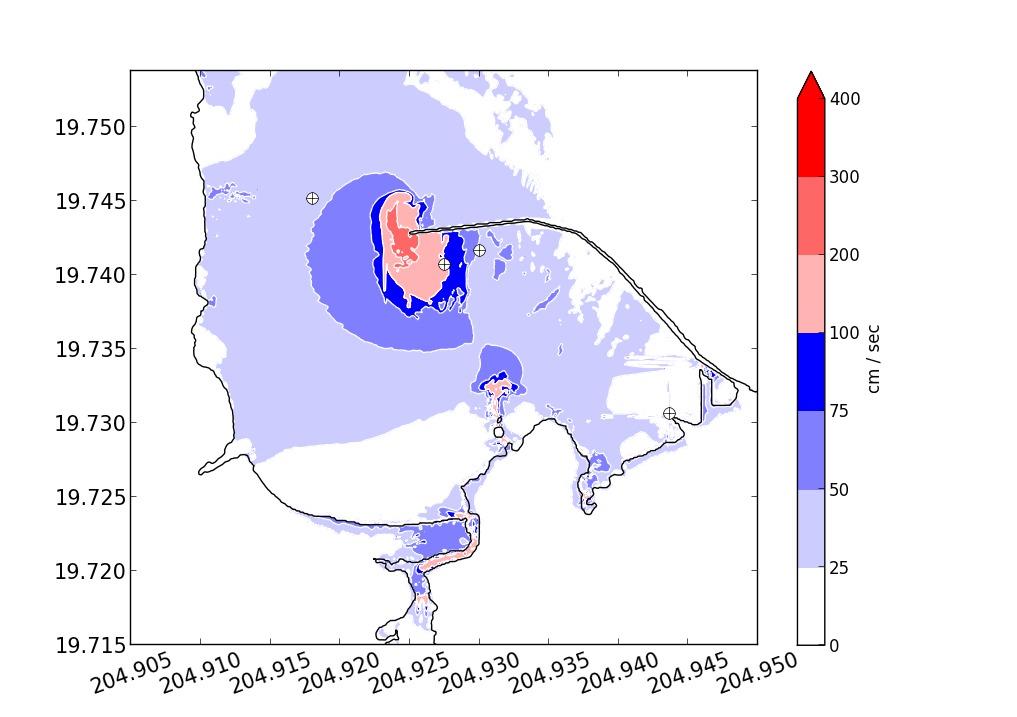}\hfil
\caption{\label{fig:Hilo} 
Left: Station locations near Hilo Harbor, including HAI1125, 1126,
tide gauge 1617760 (TG), and a synthetic gauge S2. 
Right: Maximum flow speed from model simulation (scale in cm/s).
}
\end{figure}

Stations HAI1125 and HAI1126 are in the vicinity of Hilo Harbor,
as shown in Figure~\ref{fig:Hilo} along with tide gauge 1617760.
The tsunami currents at HAI1125 are predominantly N--S, both in the
observations and the GeoClaw results, as water flows in and out of the
harbor.  On the other hand the nearby station HAI1126 is 
very close to the E--W running seawall, and the observed velocities
here are more aligned with the seawall.
At this station there is not very good agreement between the observations
and the computed velocities.  There may be several causes for this.
Note from the right side of Figure~\ref{fig:Hilo} that very high velocities are
computed near the end of the seawall.  It is also to
be expected that strong vorticity is generated as the flow goes
around this point and that water will swirl around in the harbor.  
The perpendicular change in direction between these two nearby stations is 
further indication that flow inside the harbor is likely to be turbulent.
Station HAI1126 is also in quite shallow water (12.46 m) and close
to the wall, so we might expect fairly turbulent and perhaps fully
three-dimensional fluid behavior near this
station that could be strongly affected by small scale bathymetric features.
We also note that the tide gauge in this harbor also shows the worst
agreement with the computed results (the right panel of
Figure~\ref{fig:TG}), further indication that the flow may be too complex to
be adequately modeled by the shallow water equations in this harbor.

On the other hand, note that although the computed velocities at the
location of HAI1126 are considerably smaller than the observed velocities,
there are points nearby where the velocities are comparable. 
In particular, a gauge placed at the location marked S2 in
Figure~\ref{fig:Hilo} give the velocity time series shown in
Figure~\ref{fig:S2}, which is much more similar to the observations. 
The S2 gauge is located at $(204.92753, 19.74067)$ and is roughly 280 meters
from the recorded location of HAI1126.

\section{Arrival time}\label{sec:arrival}

The simulated
tsunami generally arrived about 10 minutes before the observed tsunami,
roughly 8 hours after the earthquake.  This phase shift corresponds to a relative error
of about 2\% in the velocity of the leading wave.  
About half of this
time difference can be explained quite well by the fact that the
linearized shallow water equations are non-dispersive, meaning that the wave
speed $\sqrt{gh}$ is independent of the wavelength.  The three-dimensional
fluid dynamics is better modeled by depth-averaged equations such as
Boussinesq or Serre equations that include additional higher order
dispersive terms.  These terms are generally negligible if the wavelength of
the tsunami is very long compared to the depth of the ocean, as is often the
case for large-scale tsunamis caused by megathrust earthquakes.  However,
the Tohoku earthquake had an unusual concentration of slip on a small 
portion of the fault plane, leading to a tsunami with a relatively short
wavelength.
The dispersion relation for better models of the fluid dynamics in this
situation is often taken to be
\begin{equation}
\omega^2 = gk \tanh(kh)
\end{equation} 
(see \cite{GonzaleziKulikov:dispersion}),
where $\omega$ is the temporal frequency and $k$ is the spatial wave number,
given by $k= 2\pi / L$ in terms of the wavelength $L$.
The depth of the ocean $h$ is assumed constant in this analysis.
From this dispersion relation it can be shown that the group velocity
$d\omega / dk$ for wavenumber $k$ can be expanded as
\begin{equation}
\omega'(k) = \sqrt{gh} \left(1 - \frac 1 2 k^2 h^2 + {\cal O}(k^4h^4)\right).
\end{equation} 
Estimating an average ocean depth of 4500 m (based on the travel time
between the source region and Hawaii) and a wavelength of roughly 200 km
from ocean-scale plots of the solution, we obtain
$\frac 1 2 k^2 h^2 \approx 0.01$, which would result in about a 1\%
change in the arrival time of the wave, or about 5 minutes.  
The remaining 5 minute time shift has also been observed by other researchers
modeling results in Hawaii using dispersive Boussinesq equations
(\cite{KirbyShi:2013,Yamazaki:2012}).
At least part of it is due to the use of an instantaneous
displacement of the sea surface at the initiation time of the earthquake, as
has been used in this study and is standard practice in tsunami modeling,
rather than modeling the dynamic seafloor motion and delayed response of the
sea surface.  Additional discrepancies may
have resulted from dispersion due to the effects of ocean or tidal currents
not modeled numerically.

\section{Discussion and conclusions}\label{sec:conclusions}
Recorded observations in many locations show velocities that are not uniform in the water
column. This may bring into question the suitability of the shallow water equations
for modeling such flows. 
We believe our results show that in fact the depth averaged equations can
often be successfully used, even for locations such as gauge HAI1119 where
Figure~\ref{fig:rawdata} shows that there is significant variation with depth
and yet Figure~\ref{fig:channels1} shows good agreement of the average velocity
with the shallow water equation results.  It should also be noted that
there appears to be less variation with depth of the velocity in
Figure~\ref{fig:rawdata}  after the tsunami arrives than before, and hence the
tsunami current may be more vertically uniform than the ambient currents.
This is an important issue that deserves further study. 

The high spatial variability of the tsunami current velocities makes it
potentially much more challenging to accurately model velocities at specific
locations than to capture surface elevation. Differences in the
location and timing of maximum tsunami elevation and velocity has been
shown in other studies (e.g. \cite{Gonzalez:2009}). In view of this we believe the
agreement seen between the GeoClaw simulations and the observations
at most of the stations studied
provides significant additional validation of the model beyond what
has been achieved by past studies.  

Model results and observations differed the most at stations HAI1123 and
HAI1126, in Kahalui and Hilo Harbors, respectively. We have discussed a
number of possible reasons, including the lack of sufficiently accurate
bathymetry  and fact that
flows are expected to be much more complex and perhaps more
three-dimensional and turbulent inside harbors.  
Moreover the extreme sensitivity of the velocity to spatial location and the
uncertainty in the precise gauge locations may limit the degree to which
tsunami models can be quantitatively validated using these observations.

When Figures~\ref{fig:Hon}, \ref{fig:Kahului}, and 
\ref{fig:Hilo} are
compared, it is evident that the location of highest velocity is
highly dependent on harbor configuration. The harbors of Honolulu, Hilo and
Kahului experience maximum velocities in different settings within the
harbor. In
Honolulu Harbor, which has a broad entrance, the maximum velocities are
simulated in a broad zone near the harbor entrance and within smaller
channels. In Kahului
Harbor, which has a narrow entrance, the maximum velocities are simulated in a swath
perpendicular to the harbor entrance with lower velocities along the
edges of the harbor. In Hilo Harbor, the highest velocities are simulated at
the end of the seawall. Other studies have shown alterations to harbor
shape and bathymetry can change tsunami behavior (\cite{dengler-uslu:2011}).

An important future direction in tsunami modeling is the simulation of sediment
transport. The capability to model both tsunami erosion and deposition will
aid in hazard analysis and is also a crucial tool in helping to
reconstruct past events from tsunami deposits.
Accurate sediment
transport simulations require that numerical models produce accurate
fluid velocities and accelerations, giving additional impetus to validate codes 
against real-world data sets such as the ones used here.

The greater spatial variability in tsunami velocity than wave height
has implications for future sediment transport studies. The greater
variability in velocities is likely a result of the behavior of the
flow as it is channelized and as it flows around bathymetric highs and
structures. For example, channelized flows between islands and through
harbor entrance are high as are flows around projecting features such
as the seawall in Hilo Harbor. Wave height does not respond as strongly to channelized
flow as current velocity. This indicates that sediment is likely not uniformally
eroded at vaious water depths but erosion is concentrated in locations
of higher flow velocities. High-resolution bathymetry will be
necessary to accurately model sediment erosion for tsunami models.

This work focused on stations near harbors and those within the
channels between the islands of Maui, Molokai, and Lanai, where there is
a persistent tsunami signal due to the seiching of
water in these protected regions.  The stations not studied in this
work were in less protected settings.  We have not yet investigated
all of them, but in preliminary studies the numerical
simulations did not match the observations as well at some of the
other sites. Stronger and more erratic background currents at these
stations play a role. In addition, we believe that reflections from distant bathymetric
features may be much more important in the observed tsunami signal
at these points, and that additional refinement over a larger portion
of the ocean may be required in the future study of these stations.

All data and computer code used in this study (both the GeoClaw simulation
code and the analysis code)  is available via the
Github repository \\
$~$ \hskip 20pt {\tt https://github.com/rjleveque/tohoku2011-paper2}.\\
We hope that the data available
from these velocity meters will also be used as benchmark tests for
other tsunami simulation codes in the future.

\section{Acknowledgments}
The authors are grateful to SeanPaul La Selle for assistance in acquiring
and processing the data.
This research was supported in part by NSF Grants
DMS-0914942 and DMS-1216732, NSF RAPID Grant DMS-1137960, the Founders Term Professorship
in Applied Mathematics at the University of Washington.


\noindent
\newpage

\hfil\includegraphics[width=\textwidth]{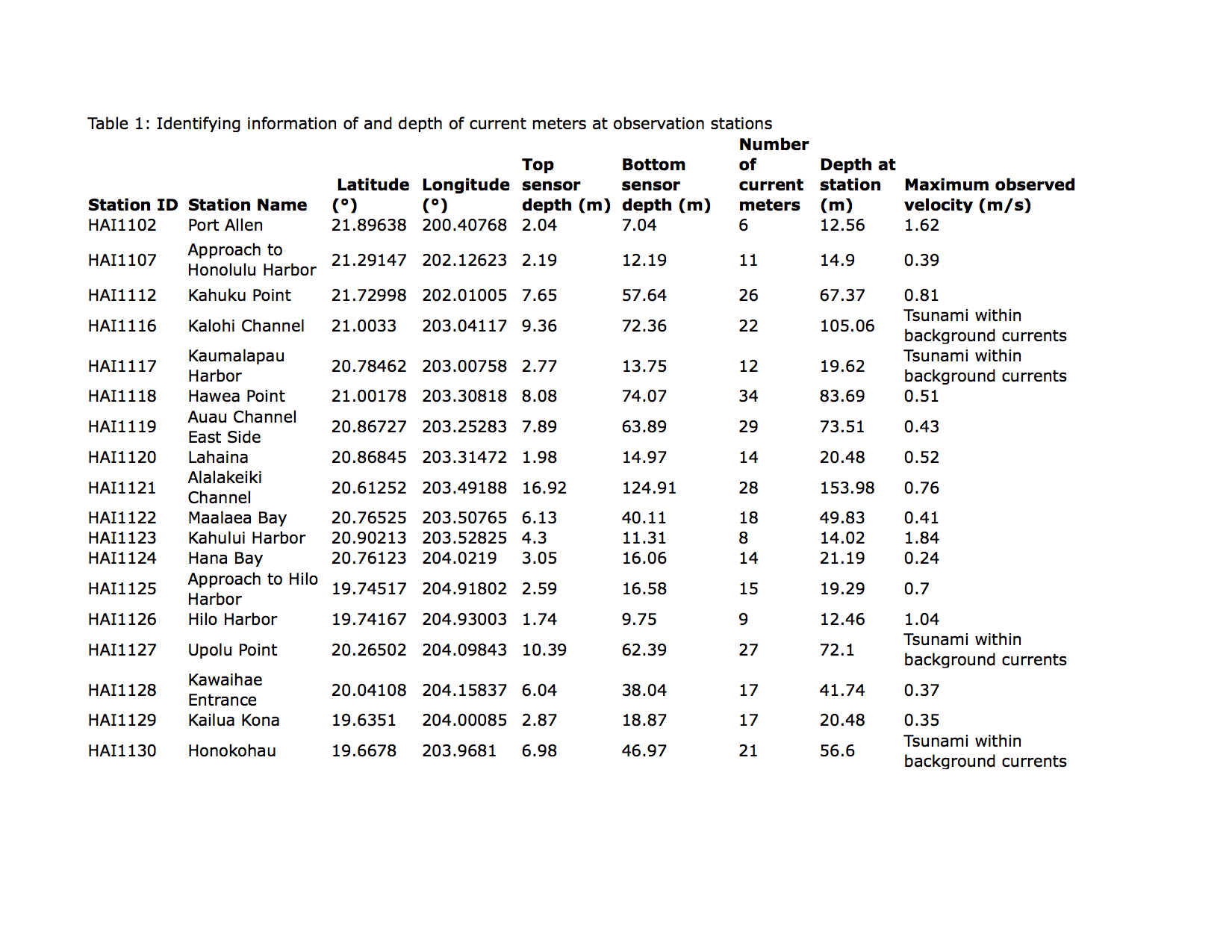}\hfil

\newpage

\begin{figure}
\hfil\includegraphics[width=0.5\textwidth]{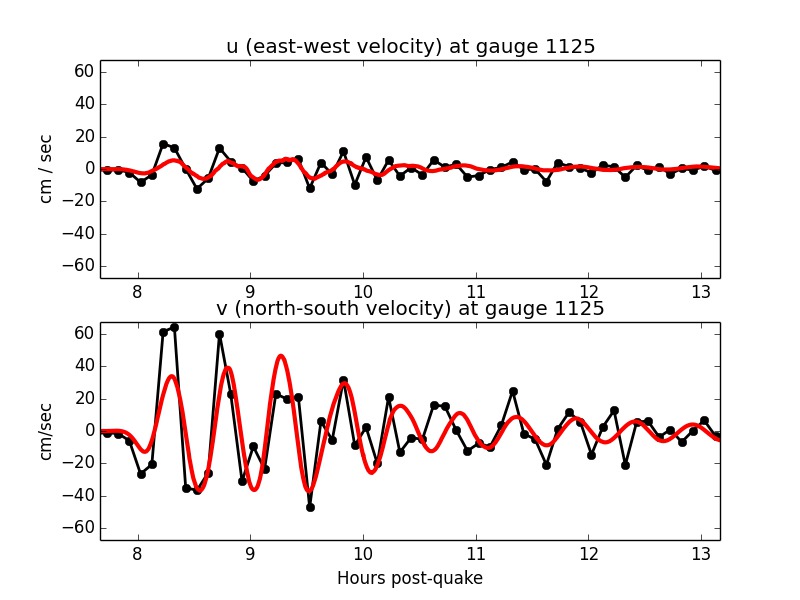}\hfil
\hfil\includegraphics[width=0.5\textwidth]{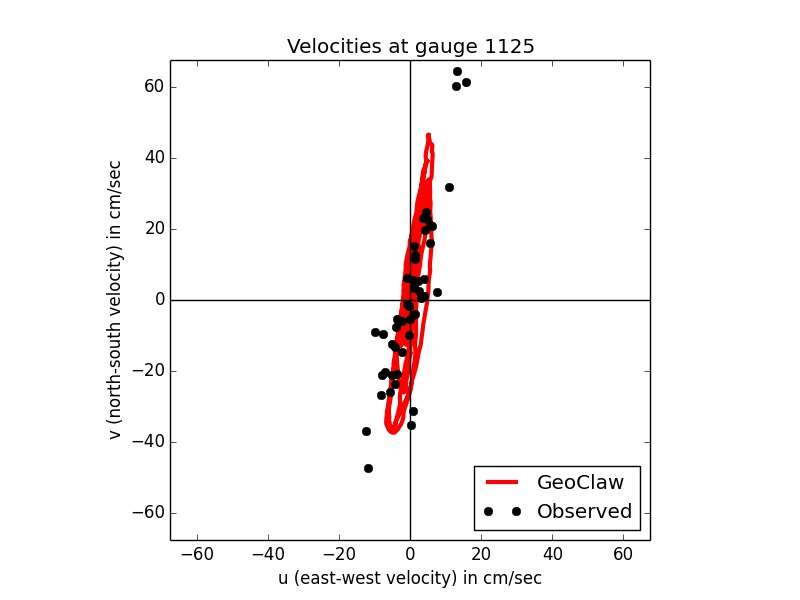}\hfil
\vskip 10pt
\hfil\includegraphics[width=0.5\textwidth]{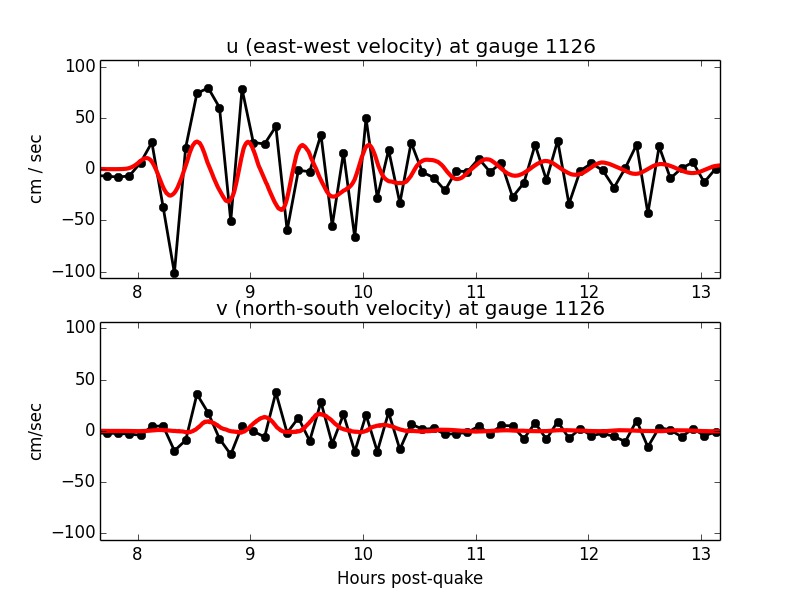}\hfil
\hfil\includegraphics[width=0.5\textwidth]{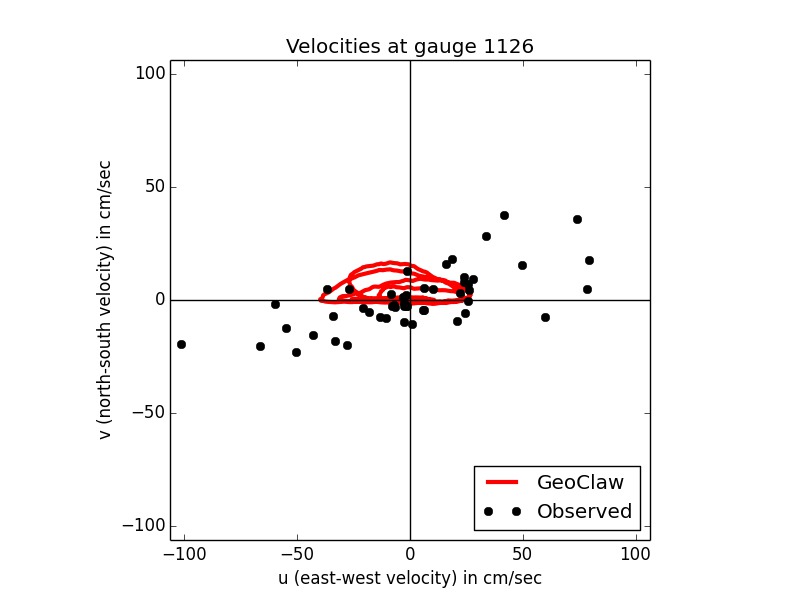}
\caption{\label{fig:HAI1125-26} 
Observed depth-averaged velocity compared to simulated
tsunami velocity at each current meter over a 6 hour window at stations
HAI1125 and 1126.  
See Figure~\ref{fig:channels1} caption for description of plots.
}
\end{figure}

\begin{figure}
\hfil\includegraphics[width=0.5\textwidth]{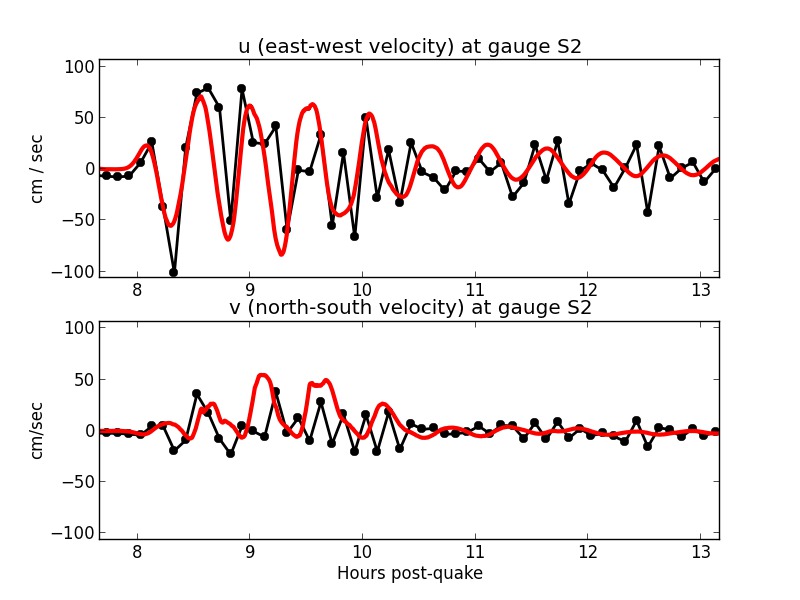}\hfil
\hfil\includegraphics[width=0.5\textwidth]{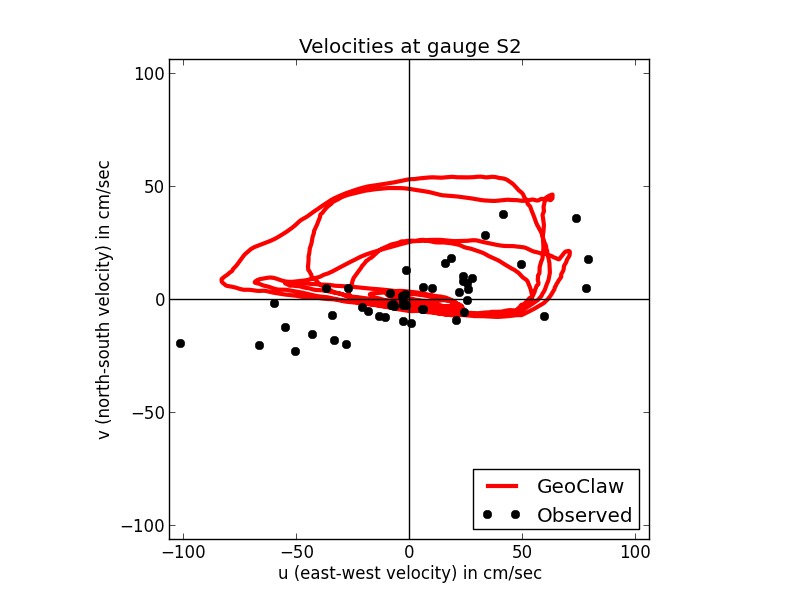}\hfil
\caption{\label{fig:S2} 
Observed depth-averaged velocity at station HAI1126 compared to simulated
tsunami velocity at synthetic gauge location S2 shown in
Figure~\ref{fig:Hilo}.
}
\end{figure}

\end{document}